\documentclass[twocolumn]{aastex631}

\begin{document}

\title[HIERACHY IV: J014741-030247]{Probing the \ion{He}{2} re-Ionization ERa via Absorbing \ion{C}{4} Historical Yield (HIERACHY) IV: A complex redshifted absorption system intrinsic to quasar}


\author[0009-0006-7138-2095]{Huiyang Mao}
\affiliation{Purple Mountain Observatory, Chinese Academy of Sciences, 10 Yuanhua Road, Nanjing 210023, PR China}
\affiliation{School of Astronomy and Space Sciences, University of Science and Technology of China, Hefei 230026, PR China}

\author[0000-0001-6239-3821]{Jiang-Tao Li}
\affiliation{Purple Mountain Observatory, Chinese Academy of Sciences, 10 Yuanhua Road, Nanjing 210023, PR China}

\author[0009-0004-1889-3043]{Xiaodi Yu}
\affiliation{Institute of Astronomy and Information, Dali University, Dali 671003, PR China}

\author[0000-0002-2941-646X]{Zhijie Qu}
\affiliation{Department of Astronomy, Tsinghua University, Beijing 100084, PR China}

\author[0000-0003-3762-7344]{Weizhe Liu}
\affiliation{Steward Observatory, University of Arizona, 933 N. Cherry Ave., Tucson, AZ 85721}

\author[0000-0001-7500-0660]{Li Ji}
\affiliation{Purple Mountain Observatory, Chinese Academy of Sciences, 10 Yuanhua Road, Nanjing 210023, PR China}

\begin{abstract}
High-resolution spectra provide a powerful tool in studying the associated absorption lines (AALs) in quasars.
We present a case study of the quasar J014741-030247 at $z \sim$ 4.75, which hosts complex intrinsic absorption lines revealed by the high-resolution Magellan/MIKE spectrum obtained from the HIERACHY program. 
We focus on one of the strongest absorption systems ($z$ $\sim$ 4.7804) and determine the column densities of multiple ionization species. We find that the Apparent Optical Depth method may significantly underestimate the column densities of high ions. Decomposing the absorption into multiple components yields a better fit and reveals clear evidence of partial coverage. The variation in covering fractions among different ions suggests that high ions are distributed more extensively in this system. We estimate electron densities of different components ($630 - 4070 \ \mathrm{cm}^{-3}$), these are based on the column densities of \ion{Si}{2}* and \ion{C}{2}*. By combining these with the hydrogen number density and ionization parameter derived from photoionization modeling, we infer that the different components are located at distances of 2.3 to 9.5 kpc from the quasar. The derived $N_{\mathrm H} / n_{\mathrm e}$ and the partial coverage observed in low ions all require cloud sizes smaller than 1 pc, even down to 0.01 pc. Finally, the low kinetic luminosity of the gas ($< 0.5\% L_\mathrm{bol}$) indicates that it is insufficient to drive significant AGN feedback and may only suppress star formation via `multistage' mechanism.

\end{abstract}


\section{Introduction}
\label{sec:intro}

Associated absorption lines (AALs) in quasars are defined as absorption features with redshifts within several thousand km~s$^{-1}$ of the quasar's emission lines (e.g., \citealt{vanden08, Hamann11, Shen12, Chen18}). These features are commonly used to probe quasar inflows and/or outflows. However, their origins can be diverse, potentially arising not only from AGN-driven gas flows but also from interstellar medium (ISM) of the host galaxy or the circumgalactic medium (CGM) on larger scales (e.g., \citealt{Foltz86, Tripp98}).

If the absorbing gas is physically associated with the quasar or its host galaxy rather than arising from unrelated intervening material along the line of sight, we consider the AALs to be intrinsic. Determining whether an AAL is intrinsic requires specific criteria, including: (1) high gas densities inferred from the ratio of excited-state to ground-state lines of the same ion; (2) detectable variability in the absorption lines over time (e.g., \citealt{Hamann97, Narayanan04}); (3) partial covering of the background source, which suggests the presence of very small absorbing clouds (e.g., \citealt{Ganguly99, Arav08}); and (4) significantly broader absorption lines than those expected from thermal or dynamical broadening in the CGM (e.g., \citealt{Misawa07}). These characteristics, often found together, strongly indicate an intrinsic origin.

Mass flux and kinetic luminosity quantify how much mass and energy AGN-driven outlfows transport, enabling assessments of their potential to drive feedback in galaxy evolution. Both the mass flux ($\dot{M}$) and kinetic luminosity ($\dot{E}_\mathrm{k}$) of the gas flow scale linearly with the total hydrogen column density ($N_\mathrm{H}$) and the distance of the gas cloud ($R$) from the central engine. Observationally, $N_H$ and $R$ can be using multiple ions, which are used to solve photoionization models and determine the ionization parameter $U$. Additionally, absorption lines produced by the excited and metastable levels of the same ion provide a density diagnostic, as their ratios are sensitive to the electron density $n_\mathrm{e}$. The gas flow distance $R$ can then be derived from the ionization parameter $U$ and the electron density $n_\mathrm{e}$. This method has been successfully applied in a series of studies (e.g., \citealt{Arav05, Arav08, Dun10, arav13, Xu18})

In low-resolution spectra, absorption troughs are typically unresolved and treated as single components using the apparent optical depth (AOD) method, which can introduce biases in the derived physical parameters of the gas flow (e.g., \citealt{Miller18, Xu18, Xu20b, Byun22b}). High-resolution rest-frame UV spectra are therefore essential for decomposing broad AALs into individual components and accurately determining their column densities and other parameters.

The ``Probing the \ion{\textbf{H}e}{2} re-\textbf{I}onization \textbf{ER}a via \textbf{A}bsorbing \ion{\textbf{C}}{4} \textbf{H}istorical \textbf{Y}ield (HIERACHY)'' program is designed to investigate \ion{He}{2} reionization and related processes at $z \sim 3-5$ using spectra against bright background quasars obtained with the 6.5m Magellan telescope \citep{Li25}. The first release of the \ion{C}{4} absorber catalog, based on high-resolution ($R \gtrsim 30,000$) spectra of 26 quasars at $z \sim 3.9-5.2$ observed with the Magellan Inamori Kyocera Echelle (MIKE) spectrograph, is published in \cite{Yu25}. Pronounced AALs are detected in about one-third of the spectra, providing an optimal dataset for studying quasar outflows and inflows during a critical epoch of cosmic reionization.

Here, we report an intriguing case of three strong AALs in the spectrum of the quasar J014741-030247 ($z_\mathrm{QSO} \sim 4.75$), all exhibiting velocity shifts of $|\Delta v| < 5000~\rm km~s^{-1}$. We focus on one of the absorption systems (S1, $z \sim 4.7804$) due to its high electron density, evident partial covering, and unusually broad line width. This case provides an excellent laboratory for a comprehensive study of the properties of AALs.

This paper is organized as follows: In \S\ref{sec: data}, we describe the Magellan/MIKE observations and data reduction procedures. In \S\ref{sec: spectrum}, we employ various methods to fit the AALs, measuring column densities, covering fractions, and other key properties. Using these measurements, we determine gas densities and photoionization conditions with the \texttt{Cloudy} code. In \S\ref{sec: discuss}, we also estimate the absorber’s distance from the quasar and compare the results obtained from the AOD method and component decomposition techniques. We finally summarize our key results in \S\ref{sec: summary}. All reported uncertainties correspond to 1~$\sigma$ (68\%) confidence levels unless otherwise stated.

\section{Magellan/MIKE Observations} \label{sec: data}

\begin{figure*}
\begin{center}
\plotone{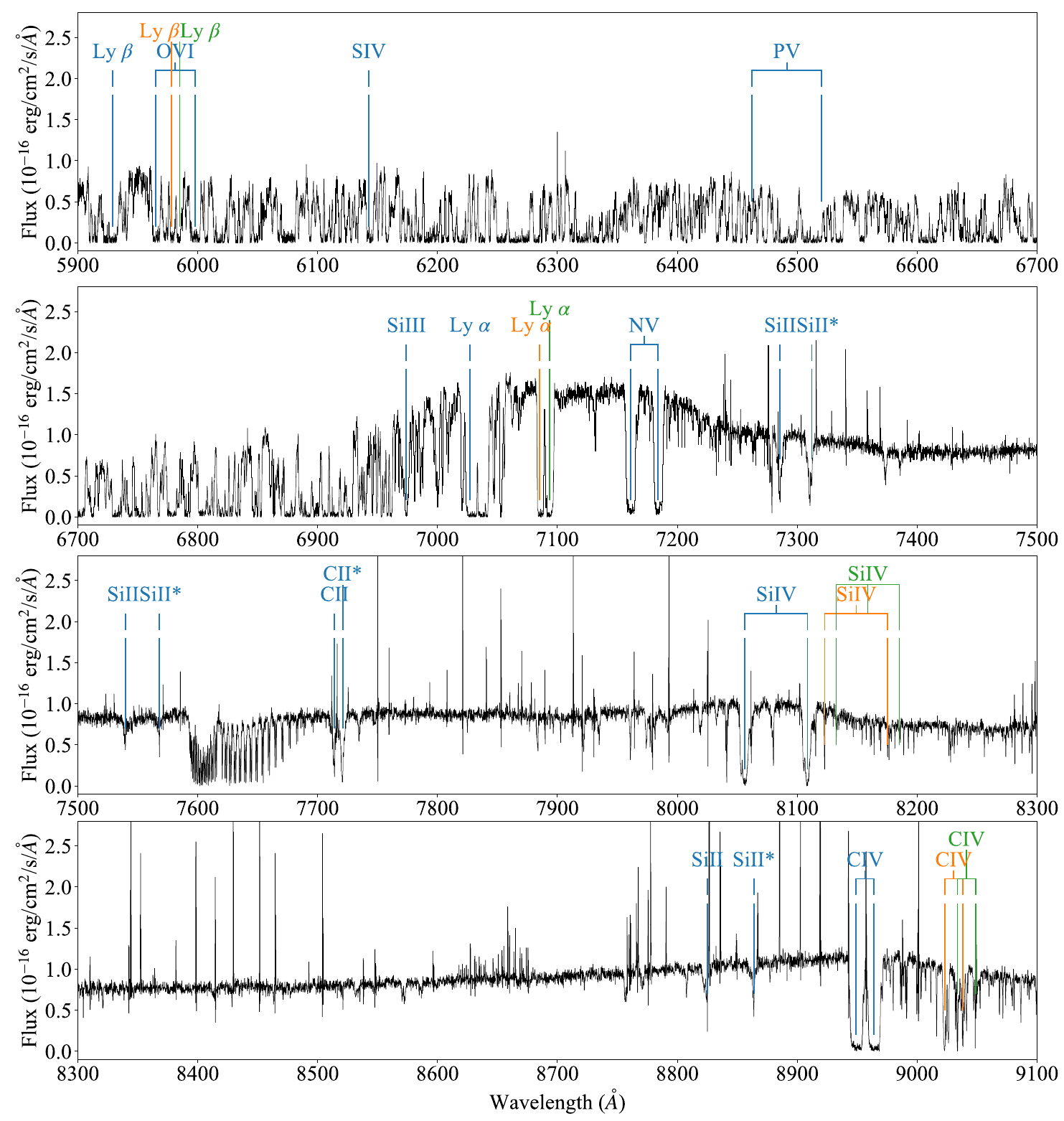}
\caption{Magellan/MIKE red channel spectrum of quasar J014741-030247 at $z \sim 4.75$. Absorption troughs of different ions are labeled with colored bars. Individual absorption lines associated with systems S1, S2, and S3 are labeled in blue, orange, and green, respectively. It is worth noting that in S2 and S3, only \ion{C}{4}, \ion{Si}{4}, Ly$\alpha$, and Ly$\beta$ lines are detected.}
\label{fig:general}
\end{center}
\end{figure*}

J014741-030247 (R.A. = 01h 47m 41.53s, Dec. = 03$^\circ$ 02' 47.88") was observed with Magellan/MIKE for $\approx 2.8~\rm hours$ ($5\times1800~\rm s$ + $1\times1000~\rm s$) on November 14, 2020, as part of the HIERACHY program \citep{Li24, Yu25}. The red channel of its global spectrum has been published in \citet{Yu25} and is also presented in Fig.~\ref{fig:general}. The redshift, $z \sim 4.75$, was estimated by fitting its optical spectrum taken with a smaller telescope with the SDSS quasar template using the ASERA toolkit \citep{Wang16}.

Detailed observation setup and data reduction procedures, including continuum determination and metal absorption line identification, are described in \citet{Li24}. In summary, we used a $0\farcs7 \times 5\farcs0$ entrance slit and applied a $2 \times 2$ binning to the detector. The spectral resolution is $R \sim 32,000$ in the red channel (4800-9400~\AA) and $R \sim 41,000$ in the blue channel (3500-5000~\AA). In this paper, we only use data from the red channel, which covers the Ly$\alpha$ emission line of the quasar and the redder portions of the spectrum that are not blended by the Ly$\alpha$ forest. The signal-to-noise ratio (S/N) of the continuum is $\sim18$ near $7100$~\AA, approximately at the Ly$\alpha$ emission line.  (Fig.~\ref{fig:general}).

\begin{figure*}[ht!]
\begin{center}
\plotone{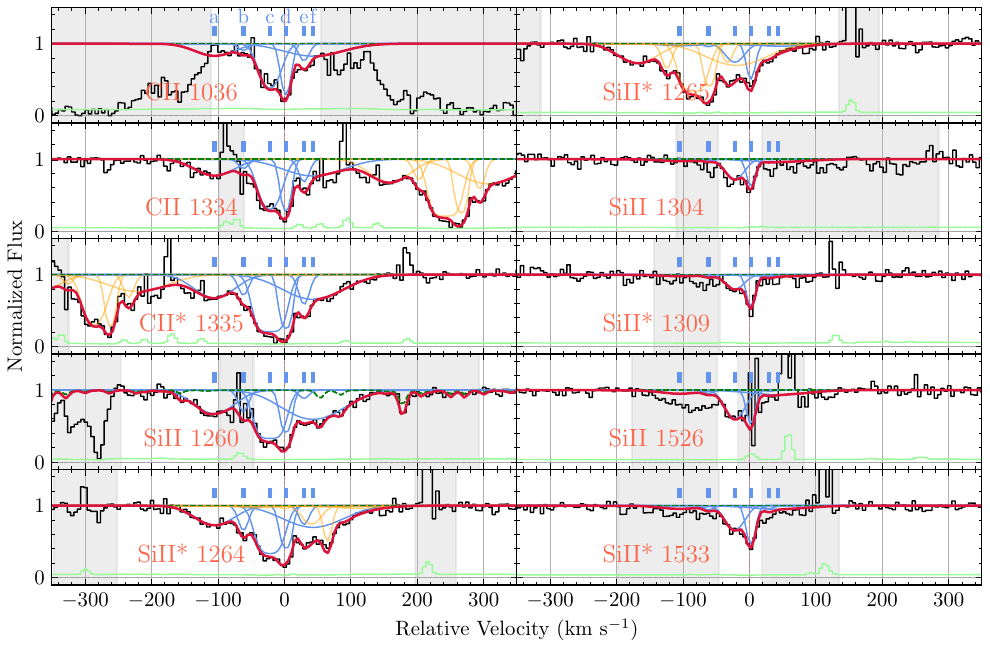}
\caption{Normalized spectrum shows the absoprtion of \ion{C}{2}, C{\small~II*}, \ion{Si}{2} and Si{\small~II*} belonging to the system S1 at the redshift of 4.7804. Six components are marked with a bar in blue, and absorption from other lines is in yellow. The absorption from the atmosphere is plotted in a dashed green line. Unrecognized absorptions and emission lines are masked with grey regions. The error of the normalized flux is plotted at the bottom of each figure in green.  
\label{fig:zoomin}}
\end{center}
\end{figure*}

\begin{figure*}[ht!]
\begin{center}
\plotone{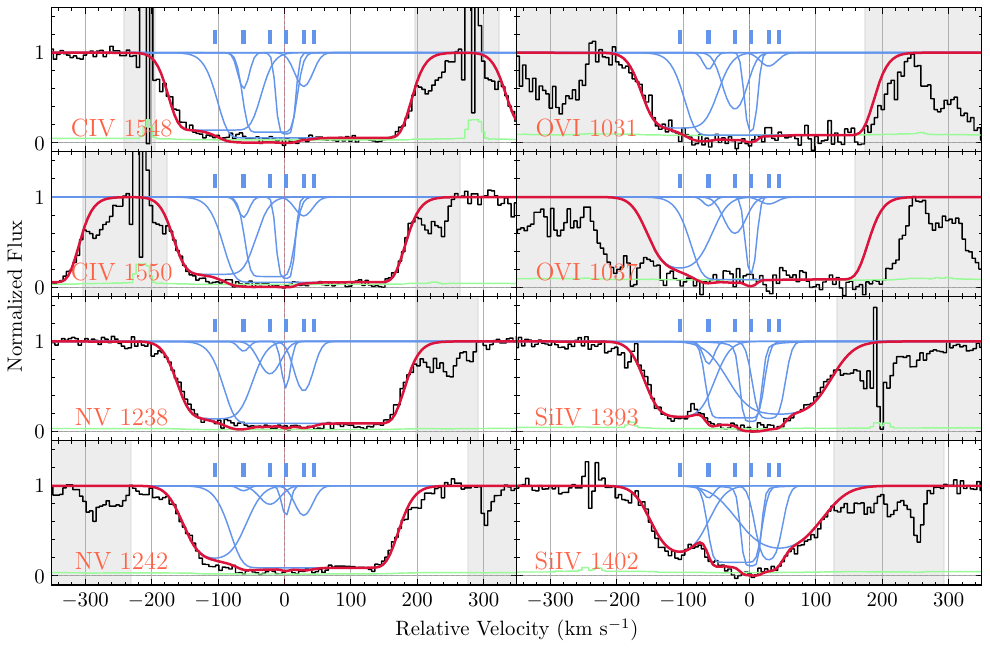}
\caption{Normalized spectrum shows the absorption of \ion{C}{4}, \ion{Si}{4}, \ion{N}{5} and \ion{O}{6} belonging to the system S1 at the redshift of 4.7804. Velocity centroids and nonthermal broadening of six components are fixed to the low ions fitted values. 
\label{fig: High_ionized}}
\end{center}
\end{figure*}

\begin{figure*}[ht!]
\begin{center}
\plotone{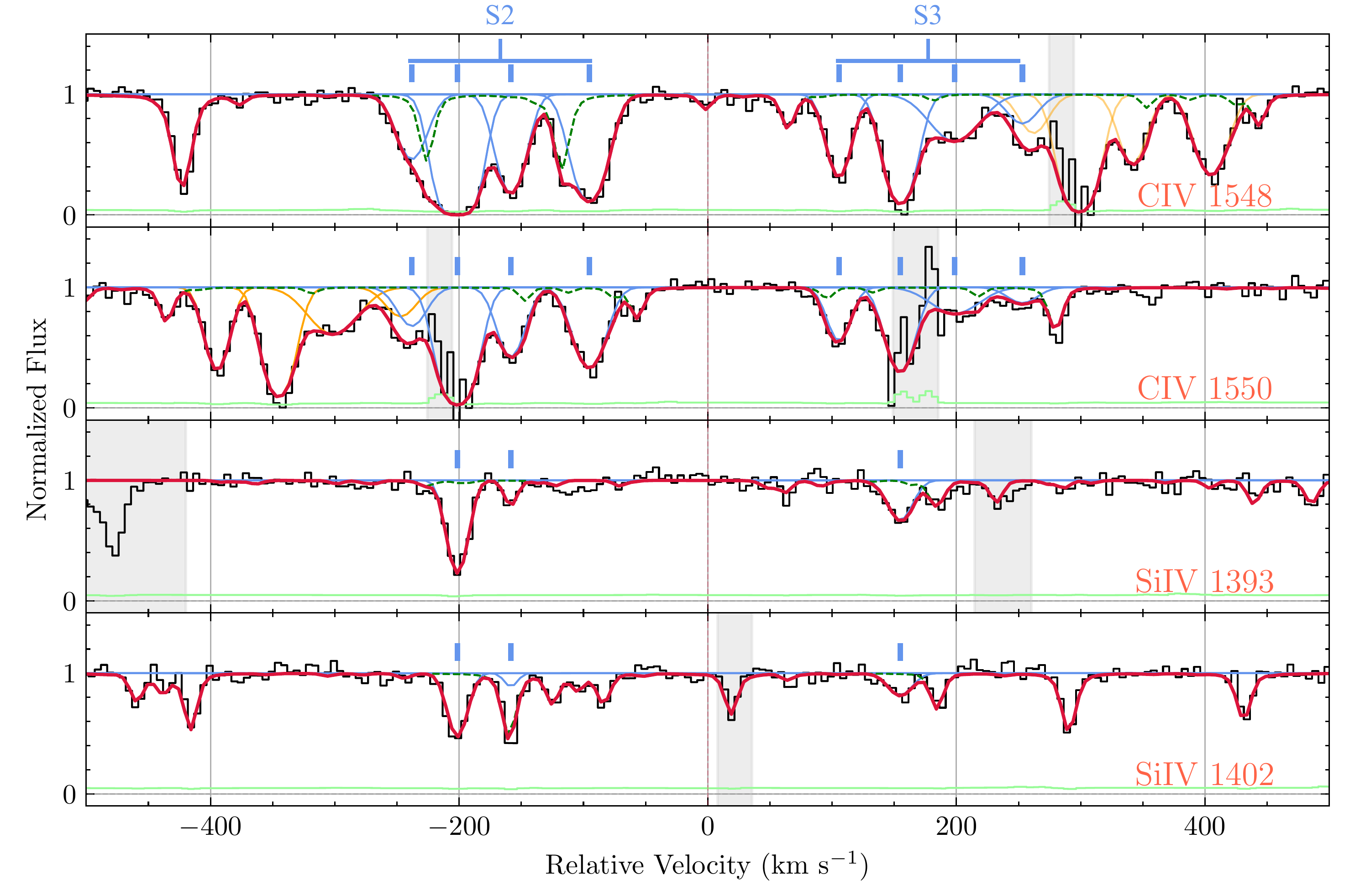}
\caption{Normalized spectrum shows the absorption of \ion{C}{4} and \ion{Si}{4} belonging to the S2 and S3 systems at the redshift $z=4.832$. The component is marked with a blue bar while its column density is larger than 10$^{12}$ cm$^{-2}$. The grey area is the region with contamination that has been masked.
\label{fig: S2columndensity}}
\end{center}
\end{figure*}

In this paper, we focus on absorption systems produced by quasar outflows or inflows, which are selected based on having velocity offsets of $|\Delta v| \leq 5000~\rm km~s^{-1}$ from the quasar's redshift. Three such absorption systems are identified and labeled as S1-S3. In the first system (S1), at $v \sim +1600~\rm km~s^{-1}$ (redshifted relative to the quasar's systemic redshift determined in \citealt{Wang16}), we detect absorption lines from multiple ions, including \ion{O}{6} $\lambda\lambda1031, 1037$, \ion{N}{5} $\lambda\lambda1238, 1242$, \ion{Si}{4} $\lambda\lambda1393, 1402$, \ion{C}{4} $\lambda\lambda1548, 1550$, as well as both ground and excited-state (denoted with $^*$) absorption lines of \ion{C}{2} and \ion{Si}{2} (Figs.~\ref{fig:zoomin} and \ref{fig: High_ionized}). In contrast, for S2 at $v \sim +4000~\rm km~s^{-1}$ and S3 at $v \sim +4400~\rm km~s^{-1}$, we only detect \ion{C}{4}, \ion{Si}{4}, Ly$\alpha$, and Ly$\beta$ absorption lines (Fig.~\ref{fig: S2columndensity}).

\section{Data Analysis} \label{sec: spectrum}

\subsection{Ionic Column Density}
\label{sec: column density}

Here we try three commonly used methods to measure the column density, and ultimately adopted the results from the apparent optical depth (AOD) method and Voigt profile methods. 

\subsubsection{AOD method}
For high ions such as \ion{C}{4}, \ion{N}{5}, and \ion{O}{6}, many of the absorption troughs are either saturated or blended with the Ly$\alpha$ forest, complicating accurate component fitting. A straightforward method to constrain the lower limit of their column densities is AOD technique (e.g., \citealt{Savage91}). The optical depth $\tau(\lambda)$ is defined in the following form (e.g., Equation (1) of \citealt{Savage91}):
\begin{equation}\label{eq:tau}
I(\lambda) = I_0(\lambda) e^{-\tau(\lambda)},
\end{equation}
where $I(\lambda)$ is the observed intensity of an absorption line, while $I_0(\lambda)$ is the intensity of the neighbouring continuum, or the intrinsic intensity before absorption. When the optical depth is expressed as a function of velocity, it relates to the column density, $N(v)$, in the following form (e.g., Equation (8) of \citealt{Savage91}):
\begin{equation}
\tau(v) = \frac{\pi e^2}{m_\mathrm{e} c} f \lambda N(v),
\label{eq:N2tau}
\end{equation}
where $m_\mathrm{e}$ is the electron mass, $e$ is the elementary charge, $f$ is the oscillator strength, and $\lambda$ is the rest-frame wavelength of the transition. Integrating $N(v)$ over the relevant velocity range yields the total column density. Because this method does not account for saturation or partial covering, it provides only a lower limit.

Using this approach, we estimate the total column densities for the strongest absorption troughs in each system. For the S1 system, we integrate over the velocity range 1400--1750~km~s$^{-1}$. For S2 and S3, the integration ranges are 4000--4200~km~s$^{-1}$ and 4360--4550~km~s$^{-1}$, respectively. The resulting column densities and associated $1\sigma$ uncertainties are presented in Table~\ref{table:ColumnDensity} and Table~\ref{table:ColumnDensity2}.

\subsubsection{Partial covering}
\label{sec: PC}

The AOD method provides a reliable estimate of the column density only when the optical depth is small ($\tau \ll 1$). However, the column densities inferred from the two lines of an ionic doublet are often inconsistent. This mismatch can be explained by the partial homogeneous covering model, in which the absorbing gas only partially covers the background source. Furthermore, although the absorption features of high ions are often nearly saturated, the residual intensity in the absorption troughs often remains nonzero (Fig.~\ref{fig: High_ionized}), indicating that complete coverage is not achieved. We therefore adopt the partial covering (PC) method, which incorporates a velocity-dependent covering fraction $C(v)$ (e.g., \citealt{Kool02, Arav05}). $C(v)$ is determined by solving the following relation:
\begin{equation}\label{eq:PartialCover}
I_n(v)-[1-C(v)]=C(v) e^{-\tau_n(v)},
\end{equation}
where $I_n(v)$ denotes the normalized intensities of the absorption line, and $\tau_n(v)$ represents the optical depth, as given in Eq. ~\ref{eq:N2tau}, which can subsequently be converted to column density. In real measurement, there are two types of observational case where the covering fraction can be determined, (1). Doublet with such as \ion{C}{4}$ \lambda\lambda1548, 1550$, (2). multiple lines from the same ion. In the first case, $n=1$ and $2$. the ratio of $\tau_1(v)$ to $\tau_2(v)$ is a constant determined by both the wavelengths and oscillator strengths. With two equations, both $C(v)$ and $\tau_1(v)$ (or $\tau_2(v)$) can be solved. In the second case, if $n$ absorption lines are observed, we obtain $n$ values of $I(v)$, and the ratios between the corresponding $\tau_n(v)$ values are constants determined by the wavelengths and oscillator strengths. In this case, the values of $\tau(v)$ and $C(v)$ can be determined numerically by minimizing the $\chi^2$ between the modeled and observed intensities in each velocity bin.

We rebin the spectra to a resolution of $20\rm~km~s^{-1}$ to improve the signal-to-noise ratio. The regions contaminated by other lines and atmospheric absorption lines are masked and excluded from the calculation. The $C(v)$ is calculated for each bin. For the doublet lines identified in the S1 system—\ion{C}{4} $\lambda\lambda1548, 1550$, \ion{Si}{4} $\lambda\lambda1393 ,1402$, \ion{N}{5} $\lambda\lambda1238, 1242$, and \ion{O}{6} $\lambda\lambda1031,1037$, these are consistent with the conditions described in the first case. So we can obtain analytical solutions for $\tau$ and $C(v)$. The same method is applied to the \ion{C}{2} case (\ion{C}{2} $\lambda 1036$, \ion{C}{2} $\lambda 1334$), which aligns with the second case. For \ion{Si}{2}, We have identified numerous lines, which correspond to the third case, meaning we can only obtain a numerical solution. In principle, the covering fractions for different states of the same ion, such as \ion{Si}{2}* and \ion{Si}{2}, should be consistent. However, the potential difference in column density may cause the ratio between their $\tau$ values to not be constant. Therefore, here we calculate $C(v)$ separately for \ion{Si}{2} and \ion{Si}{2}* (for \ion{Si}{2}, \ion{Si}{2} $\lambda1260$, \ion{Si}{2} $\lambda1304$, \ion{Si}{2} $\lambda1526$; for \ion{Si}{2}*, \ion{Si}{2}* $\lambda1264$, \ion{Si}{2}* $\lambda1265$, \ion{Si}{2}* $\lambda1309$, and \ion{Si}{2}* $\lambda1533$) to derive the covering fractions. For S2 and S3, atmosphere absorption significantly overlaps with the ionic absorption features (indicated by the dashed green lines in Fig. \ref{fig: S2columndensity}), and the \ion{C}{4} doublets are blended with each other. Minor deviations in the doublet ratio can result in substantial inaccuracies in the determination of both the covering fraction and the column density. Thus PC method is not applied to the S2 and S3 systems.

\begin{figure*}[ht!]
\begin{center}

\includegraphics[width=2.0\columnwidth]{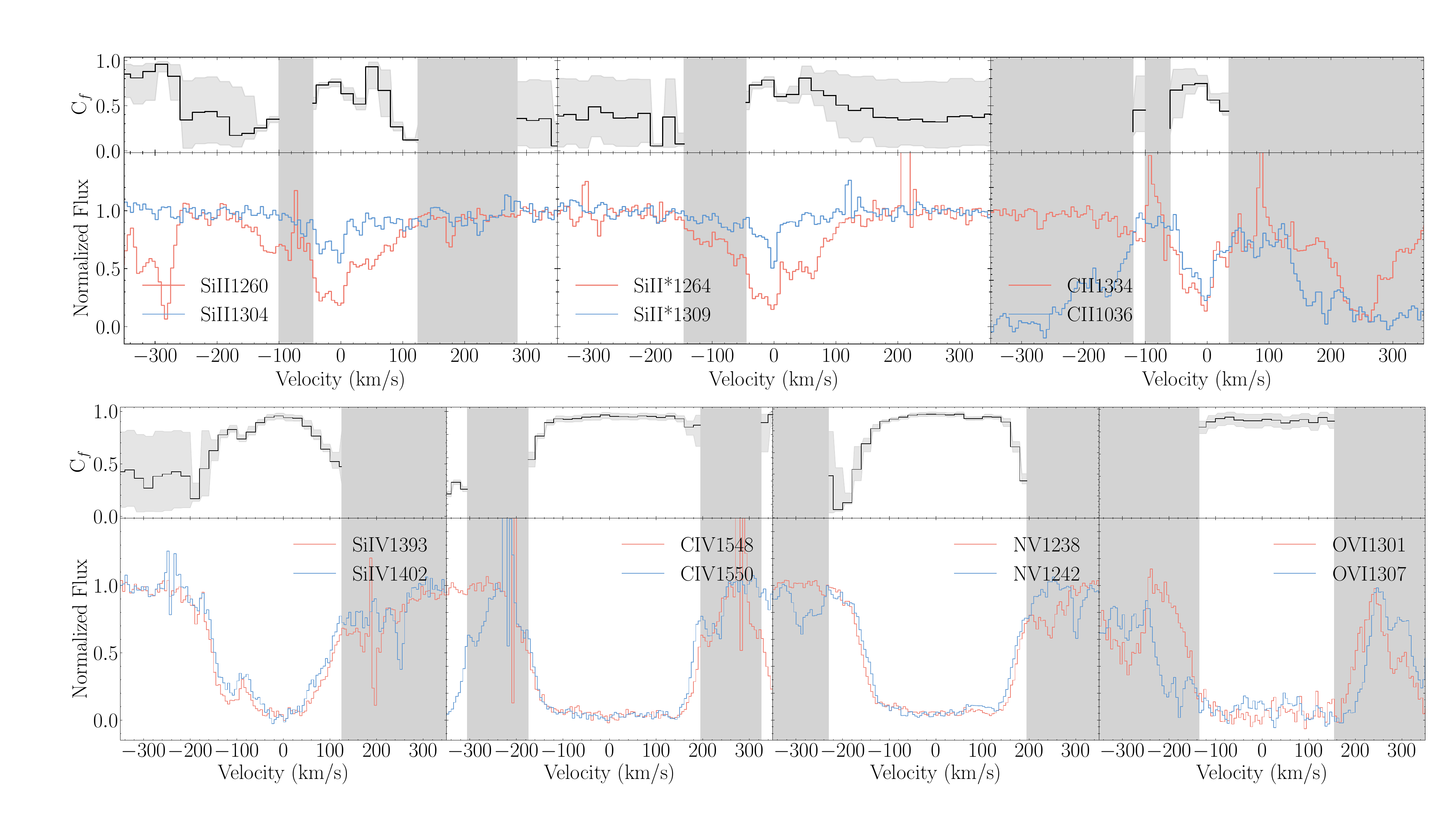}
\caption{Partial covering fitting results of different ion lines. The covering fraction of each bin is plotted on top of the normalized spectrum. Grey region represents the range of one sigma error. Absorption contaminated is masked by grey rectangles and the covering fractions of these bins are not shown in the figure. Different line profiles of the same ions are plotted in the same velocity space.
\label{fig:partial_covering}}
\end{center}
\end{figure*}

Fig. \ref{fig:partial_covering} shows the covering fraction calculated for each ion. Although the uncertainties are large and it is difficult to obtain reliable column densities, we can still see from the figure that, within the error range, many bins for each ion exhibit covering fractions significantly less than 1.

\subsubsection{Decomposition Fitting}

The absorption of low ion clearly calls for multiple components, and their column densities and b factors appear to differ significantly. Therefore, decomposition fitting is necessary. High-resolution MIKE optical spectra allow for the decomposition of each absorption feature into multiple components and enable precise comparison of velocity alignment across different ionic species (e.g., \citealt{Cooper21, Zahedy21, Qu22}). Therefore, we use Voigt profile fitting for multiple components to obtain column densities here. The Voigt profile fitting allows for the inclusion of atmospheric line models, thereby minimizing the impact of telluric absorption features on the results. Our principle for characterizing each complex multi-component system is to use the minimum number of components whenever possible.

We start identifying different components with the most complex S1 system. The lines from high ions (\ion{O}{6}, \ion{N}{5}, and \ion{C}{4}) in S1 are strongly saturated, making it difficult to distinguish individual absorption components from these transitions alone. Fortunately, several lower-ionization lines—particularly multiple transitions of \ion{C}{2} and \ion{Si}{2}, including their excited states—are unsaturated and offer better constraints for decomposing the system. Based on these transitions, we identify six distinct absorption components in S1. Among them, components $\mathbf{c}$ and $\mathbf{d}$ are the most prominent, with strong absorption signatures consistently detected across all available transitions. Components $\mathbf{a}$ and $\mathbf{f}$ appear as broad, shallow troughs in \ion{C}{2} $\lambda1334$, \ion{C}{2}* $\lambda1335$, \ion{Si}{2} $\lambda1260$, and \ion{Si}{2}* $\lambda1264$. Component $\mathbf{b}$ is relatively weak and is primarily identified through the excited-state transitions \ion{C}{2}* $\lambda1335$ and \ion{Si}{2}* $\lambda1264$. Finally, component $\mathbf{e}$ is clearly distinguished in most detected lines.

For comparison, the S2 and S3 systems are simpler with less absorption components from various ions. The most prominent absorption lines in S2 and S3 are \ion{C}{4} and \ion{Si}{4}, with the two systems separated by $\sim400~\rm km~s^{-1}$ (Fig.~\ref{fig: S2columndensity}). We use the stronger \ion{C}{4} $\lambda1548$ line as the primary tracer for identifying different absorption components. This line is free from contamination by other metal lines, although some telluric features are present in the same wavelength range and are indicated by green dashed curves in Fig.~\ref{fig: S2columndensity}. In total, four distinct absorption components are resolved in S2. For S3, although the \ion{C}{4} $\lambda1548$ line is blended with the $\lambda1550$ line from S2, four components can still be clearly identified.

We attempt to simultaneously fit all low ions, with their velocity centroids linked across components. The total normalized intensity can be written as the product of the intensities after absorption of each component:

\begin{equation}
I(v)=\prod\left\{C_f e^{-\tau(v)} + 1 - C_f \right\},
\end{equation}

where $C_f$ and $\tau(v)$ represent the covering fraction and optical depth respectively, and the optical depth can be expressed as:
\begin{equation}
\tau(v \mid N,b,z)=N \sigma_0 f \Phi(v \mid b,z),
\end{equation}

where $N, b, z, f, v$ and $\sigma_0$ are the ion column density, Doppler parameter, the redshift of the system, oscillator strength,  cross-section of the ion, and the velocity relative to the system. $\Phi(v \mid b,z)$ is the normalized Voigt profile. In the same system, the velocity centroid of the same component from different ions is linked with each other. 

Doppler broadening is caused by thermal and nonthermal motions (e.g., \citealt{Rauch96}), $b^2=b_\mathrm{T}^2+b_{\mathrm{NT}}^2$. There are currently no other reliable temperature-sensitive indicators available for $b_\mathrm{T}$. So we assume a temperature of $10^4$ K to fix its value. The typical temperature range in which \ion{C}{2} and \ion{Si}{2} exist is between $10^4$ K and $10^5$ K (e.g., \citealt{Gnat07}). Within this temperature range, the variation of the $b_T$ ($\sqrt{2k_\mathrm{BT}/m_\mathrm{i}}$) is relatively small ($< 8$ km/s). Then we link the nonthermal line width $b_\mathrm{NT}$ of the components from \ion{C}{2} with the one from \ion{Si}{2}. 

For high ions, since the absorption troughs of ions other than \ion{Si}{4} are heavily saturated and do not allow clear separation of different components, we tie the velocity centroids of each component to those derived from the \ion{Si}{2} and \ion{C}{2} fits. High ions (\ion{Si}{4}, \ion{C}{4}, \ion{N}{5}, \ion{O}{6}) typically exist at a temperature of around $10^5$ K (e.g., \citealt{Gnat07}). Therefore, the thermal broadening between these ions should not differ significantly; instead, the nonthermal broadening, which is the same among these ions, should dominate. Saturation also makes it difficult to determine the precise $b_{\mathrm{NT}}$ for each component; therefore, we set the $b_{\mathrm{NT}}$ of all components to the same values as the those of \ion{Si}{2} and \ion{C}{2}, which may underestimate the $b_{\mathrm{NT}}$ of high ions. The column densities and covering fractions of different components for various ions are set to be free for fitting.

In Sec. \ref{sec: PC}, we find that the covering fraction in S1 is significantly less than 1, so we also take the covering fraction into account in the Voigt profile fitting. We obtain covering fractions of low ions in two cases. For components with sufficiently strong absorption (component \textbf{c} and \textbf{d}), we treat the covering fraction as a free parameter in the fitting. For the remaining components, we used the covering fractions of high ions as upper limits, they exhibit strong absorption in high ions, allowing their fitted results to be used as upper limits for low ions. We find that components \textbf{a} and \textbf{f} show nearly identical profiles in the high ions doublets, indicating clear evidence of partial covering. Therefore, we treated the covering fractions of components \textbf{a, c, d}, and \textbf{f} as free parameters, while fixing those of components \textbf{b} and \textbf{e} to 1. Among the four ions (\ion{Si}{4}, \ion{C}{4}, \ion{N}{5}, \ion{O}{6}), component \textbf{a} of \ion{N}{5} has the lowest covering fraction, 0.78, and component \textbf{f} of \ion{Si}{4} has the lowest covering fraction, 0.80. We adopt these values as the upper limits for the covering fractions of components \textbf{a} and \textbf{f} of \ion{C}{2} and \ion{Si}{2}, respectively. The lower limits of the partial covering are obtained from the lowest normalized flux from the components. The results are presented as a range and shown in Table \ref{table: result}.

From Sec. \ref{sec: PC}, we cannot confirm the presence of significant partial covering in S2 and S3; therefore, the covering fractions are fixed to 1. The results of covering fractions are shown in Table \ref{table:ColumnDensity}. In the subsequent fitting and analysis, we adopt the column densities derived from the upper limits of the covering fractions. 

\begin{figure}
\includegraphics[width=1.0\columnwidth]{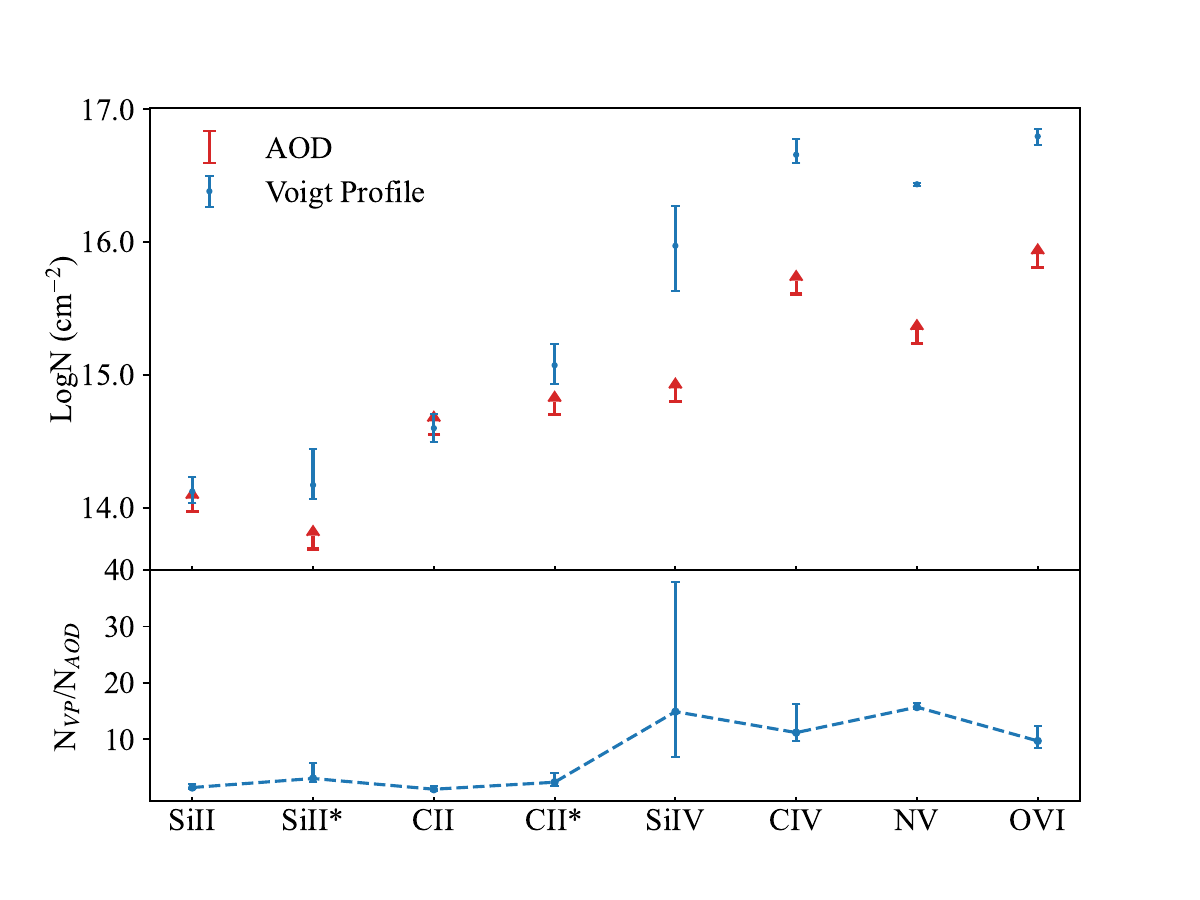}
\caption{The upper panel is the total column density of each ion measured by AOD and Voigt profile fitting. Blue dots with error bars represent the sum of column densities of six components measured by the Voight profile and the red one shows the lower limit measured by AOD. The lower panel shows the ratio between the column densities measured by Voigt profile fitting and AOD.
\label{fig: col_compare}}
\end{figure}

\begin{figure}
\includegraphics[width=1.0\columnwidth]{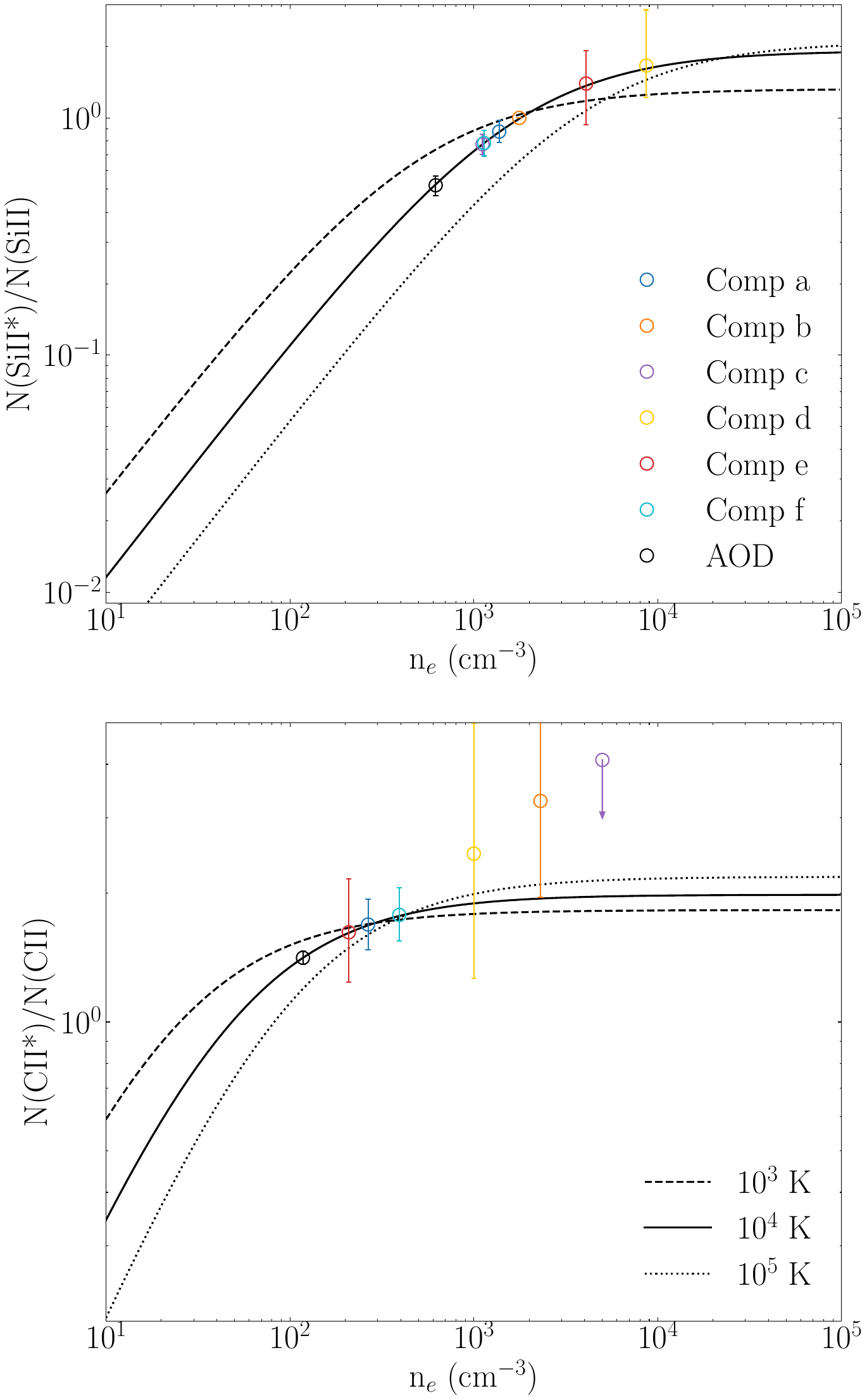}
\caption{Upper panel: \ion{Si}{2}*/\ion{Si}{2} column density ratio vs. $n_e$. The black lines with different line styles represent the same relationships under different temperatures. Colored circles with one sigma error bar represent density solutions from different components. Lower panel: \ion{C}{2}*/\ion{C}{2} column density ratio vs. $n_e$, the other are the same as the upper panel.
\label{fig: SiII_ratio}}
\end{figure}

\begin{table*}
\begin{center}
\hspace{-1in}
\tabcolsep=3.5pt%
\caption{The fitting results of low ions in S1. The fitting velocity centroid is relative to $z=4.7804$. For the same component, the velocity centroid and $b_\mathrm{NT}$ are set to be identical for different ions, so the errors are only shown in \ion{C}{2} columns. The covering fraction of the same component for the same ion is assumed to be identical.  The upper limits and lower limits of covering fractions for component \textbf{a} and \textbf{f} are obtained from high ions and normalized flux seperately.}
\begin{tabular}{lccccc}
\hline\hline
Ion & logN (cm$^{-2}$) & logN (cm$^{-2}$) (AOD) & $b_{\mathrm{NT}}$ (km s$^{-1}$) & velocity centroid (km s$^{-1}$) & covering fraction  \\
\hline
\multicolumn{6}{c}{S1} \\
\hline
\ion{C}{2} 
& 13.61$^{+0.07}_{-0.06}$ & $\textgreater$ 14.55$\pm$0.03 & 39.5$^{+4.4}_{-3.7}$ & -105.4$^{+3.7}_{-3.0}$ & [0.36, 0.78] \\
& 12.76$^{+0.25}_{-0.63}$ & & 8.2$^{+2.6}_{-2.6}$ & -61.7$^{+1.4}_{-1.4}$ & 1.0 \\
& 14.20$^{+0.08}_{-0.08}$ & & 17.6$^{+2.4}_{-2.6}$ & -21.9$^{+2.2}_{-2.3}$ & 0.79$^{+0.04}_{-0.04}$ \\
& 14.27$^{+0.43}_{-0.23}$ & & 5.2$^{+1.3}_{-2.3}$ & 2.1$^{+0.5}_{-0.7}$ & 0.75$^{+0.09}_{-0.09}$ \\
& 13.20$^{+0.10}_{-0.12}$ & & 8.1$^{+2.2}_{-2.0}$ & 29.5$^{+1.2}_{-1.3}$ & 1.0 \\
& 13.79$^{+0.06}_{-0.06}$ & & 59.4$^{+3.4}_{-3.5}$ & 43.1$^{+3.6}_{-2.3}$ & [0.37, 0.80] \\

\ion{C}{2}* 
& 13.83$^{+0.04}_{-0.04}$ & $\textgreater$ 14.70$\pm$0.02 & 39.5 & -105.4 & [0.36, 0.78] \\
& 13.20$^{+0.14}_{-0.17}$ & & 8.2 & -61.7 & 1.0 \\
& 14.82$^{+0.13}_{-0.12}$ & & 17.6 & -21.9 & 0.79 \\
& 14.78$^{+0.63}_{-0.43}$ & & 5.2 & 2.1 & 0.75 \\
& 13.40$^{+0.09}_{-0.10}$ & & 8.1 & 29.5 & 1.0 \\
& 14.04$^{+0.04}_{-0.04}$ & & 59.4 & 43.1 & [0.37, 0.80] \\
 
\ion{Si}{2} 
& 12.84$^{+0.06}_{-0.05}$ & $\textgreater$ 13.97$\pm$0.04 & 39.5 & -105.4 & [0.37, 0.78] \\
& 12.21$^{+0.18}_{-0.15}$ & & 8.2 & -61.7 & 1.0 \\
& 13.74$^{+0.06}_{-0.07}$ & & 17.6 & -21.9 & 0.67$^{+0.02}_{-0.02}$ \\
& 13.70$^{+0.16}_{-0.14}$ & & 5.2 & 2.1 & 0.60$^{+0.07}_{-0.06}$ \\
& 11.74$^{+0.24}_{-0.36}$ & & 8.1 & 29.5 & 1.0 \\
& 13.13$^{+0.03}_{-0.04}$ & & 59.4 & 43.1 & [0.52, 0.80] \\

 \ion{Si}{2}* 
& 12.78$^{+0.05}_{-0.05}$ & $\textgreater$ 13.69$\pm$0.01 & 39.5 & -105.4 & [0.37, 0.78] \\
& 12.33$^{+0.10}_{-0.12}$ & & 8.2 & -61.7 & 1.0 \\
& 13.63$^{+0.06}_{-0.07}$ & & 17.6 & -21.9 & 0.67 \\
& 13.92$^{+0.23}_{-0.13}$ & & 5.2 & 2.1 & 0.60 \\
& 12.18$^{+0.13}_{-0.18}$ & & 8.1 & 29.5 & 1.0 \\
& 13.02$^{+0.05}_{-0.05}$ & & 59.4 & 43.1 & [0.52, 0.80] \\

\hline \hline

\end{tabular}
\label{table:ColumnDensity}

\footnotesize
\end{center}
\end{table*}

\begin{table*}
\centering
\caption{The fitting results of high ions in S1-S3. The velocity centroids fitting in S2 and S3 are set to be relative to $z=4.832$.}
\hspace{-1in}
\tabcolsep=3.5pt%
\begin{tabular}{lccccc}
\hline\hline
Ion & logN (cm$^{-2}$) & logN (cm$^{-2}$) (AOD) & $b_{\mathrm{NT}}$ (km s$^{-1}$) & velocity centroid (km s$^{-1}$) & covering fraction \\
\hline
\multicolumn{6}{c}{S1}\\
\hline

\ion{Si}{4} 
& 14.32$^{+0.02}_{-0.02}$ & $\textgreater$ 14.80$\pm$0.01 & 39.5 & -105.4 & 0.80$^{+0.01}_{-0.01}$ \\
& 12.41$^{+0.18}_{-0.16}$ & & 8.2 & -61.7 & 1.0 \\
& 15.87$^{+0.16}_{-0.18}$ & & 17.6 & -21.9 & 0.80$^{+0.02}_{-0.02}$ \\
& 16.14$^{+0.19}_{-0.36}$ & & 5.2 & 2.1 & 0.84$^{+0.05}_{-0.05}$ \\
& 13.95$^{+0.25}_{-0.25}$ & & 8.1 & 29.5 & 1.0 \\
& 14.50$^{+0.02}_{-0.02}$ & & 59.4 & 43.1 & 0.80$^{+0.01}_{-0.01}$ \\

\ion{C}{4} 
& 15.38$^{+0.02}_{-0.02}$ & $\textgreater$ 15.61$\pm$0.01& 39.5 & -105.4 & 0.84$^{+0.01}_{-0.01}$ \\
& 13.05$^{+0.28}_{-0.25}$ & & 8.2 & -61.7 & 1.0 \\
& 13.89$^{+0.59}_{-0.37}$ & & 17.6 & -21.9 & 0.89$^{+0.04}_{-0.04}$ \\
& 14.44$^{+0.73}_{-0.58}$ & & 5.2 & 2.1 & 0.89$^{+0.04}_{-0.03}$ \\
& 13.27$^{+0.29}_{-0.27}$ & & 8.1 & 29.5 & 1.0 \\
& 16.74$^{+0.08}_{-0.12}$ & & 59.4 & 43.1 & 0.94$^{+0.01}_{-0.01}$ \\

\ion{N}{5} 
& 15.07$^{+0.01}_{-0.01}$ & $\textgreater$ 15.24$\pm$0.01 & 39.5 & -105.4 & 0.78$^{+0.01}_{-0.01}$ \\
& 12.56$^{+0.25}_{-0.22}$ & & 8.2 & -61.7 & 1.0 \\
& 13.72$^{+0.19}_{-0.24}$ & & 17.6 & -21.9 & 0.83$^{+0.06}_{-0.05}$ \\
& 13.55$^{+0.29}_{-0.30}$ & & 5.2 & 2.1 & 0.82$^{+0.06}_{-0.05}$ \\
& 13.76$^{+0.13}_{-0.16}$ & & 8.1 & 29.5 & 1.0 \\
& 16.82$^{+0.01}_{-0.01}$ & & 59.4 & 43.1 & 0.90$^{+0.01}_{-0.01}$ \\

 \ion{O}{6} 
& 15.23$^{+0.06}_{-0.05}$ & $\textgreater$ 15.81$\pm$0.01 & 39.5 & -105.4 & 0.79$^{+0.03}_{-0.03}$ \\
& 13.06$^{+0.23}_{-0.23}$ & & 8.2 & -61.7 & 1.0 \\
& 14.08$^{+0.36}_{-0.24}$ & & 17.6 & -21.9 & 0.89$^{+0.04}_{-0.03}$ \\
& 14.45$^{+0.58}_{-0.41}$ & & 5.2 & 2.1 & 0.89$^{+0.04}_{-0.04}$ \\
& 13.14$^{+0.25}_{-0.25}$ & & 8.1 & 29.5 & 1.0 \\
& 16.92$^{+0.03}_{-0.04}$ & & 59.4 & 43.1 & 0.91$^{+0.01}_{-0.01}$ \\

\hline
\multicolumn{6}{c}{S2}\\
\hline

\ion{C}{4} 
& 13.44$^{+0.03}_{-0.02}$ & $\textgreater$ 14.43$\pm$0.02 & 14.6$^{+0.5}_{-0.5}$ & -238.0$^{+0.9}_{-0.8}$ & 1.0 \\
& 14.44$^{+0.03}_{-0.03}$ & & 14.9$^{+0.3}_{-0.3}$ & -201.4$^{+0.2}_{-0.2}$ & 1.0 \\
& 13.74$^{+0.01}_{-0.01}$ & & 12.9$^{+0.2}_{-0.2}$ & -158.4$^{+0.3}_{-0.3}$ & 1.0 \\
& 13.91$^{+0.01}_{-0.01}$ & & 15.0$^{+0.3}_{-0.2}$ & -95.2$^{+0.2}_{-0.2}$ & 1.0 \\

\ion{Si}{4} 
& \textless 12.0 & $\textgreater$ 13.19$\pm$0.02 & 14.6 & -238.0 & 1.0 \\
& 13.17$^{+0.01}_{-0.01}$ & & 14.9 & -201.4 & 1.0 \\
& 12.21$^{+0.04}_{-0.04}$ & & 12.9 & -158.4 & 1.0 \\
& \textless 12.0 & & 15.0 & -95.2 & 1.0 \\

\hline
\multicolumn{6}{c}{S3}\\
\hline
\ion{C}{4} 
& 13.52$^{+0.01}_{-0.01}$ & $\textgreater$ 14.21$\pm$0.02 & 10.9$^{+0.3}_{-0.3}$ & 105.5$^{+0.2}_{-0.3}$ & 1.0 \\
& 13.90$^{+0.02}_{-0.02}$ & & 13.4$^{+0.2}_{-0.2}$ & 154.8$^{+0.2}_{-0.3}$ & 1.0 \\
& 13.46$^{+0.02}_{-0.02}$ & & 25.3$^{+1.2}_{-1.1}$ & 198.7$^{+0.9}_{-0.9}$ & 1.0 \\
& 13.02$^{+0.04}_{-0.04}$ & & 15.6$^{+0.6}_{-0.6}$ & 253.1$^{+1.0}_{-1.1}$ & 1.0 \\

\ion{Si}{4} 
& \textless 12.0 & $\textgreater$ 12.80$\pm$0.04 & 10.9 & 105.5 & 1.0 \\
& 12.70$^{+0.02}_{-0.02}$ & & 13.4 & 154.8 & 1.0 \\
& \textless 12.0 & & 25.3 & 198.7 & 1.0 \\
& \textless 12.0 & & 15.6 & 253.1 & 1.0 \\
\hline \hline

\end{tabular}
\label{table:ColumnDensity2}

\footnotesize
\end{table*}

\subsubsection{Column densitiy comparison}

After deriving the column densities using both the AOD method and the decomposition fitting, we compare the total column density of each ion obtained from the two methods. The results are presented in Fig. \ref{fig: col_compare}. The different ions are arranged in order of increasing ionization potential, from low to high. The ratio of measured column densities between these two methods is also shown in the lower panel. The difference between measurements from the two methods is not large in the case of unsaturated lines, even though we identify significant partial covering in the absorption of \ion{Si}{2} and \ion{C}{2}, the measurements are of the same order. However, the column densities measured by Voigt profile fitting are more than 10 times those measured by AOD method in the case of high ions where the absorptions are nearly saturated. This situation is very common in broad absorption line (BAL) quasars, where absorption troughs from high ions are broad and saturated while the absorptions from low ions are moderate. Our case shows that the column densities of high ions will be significantly underestimated if measured using the AOD method, which may lead to the underestimation of the ionization parameter of the gas.

\subsection{Electron Number Density}
\label{sec: density}

We can determine the electron number density of each component by measuring the column density ratios between excited and ground states. 

\begin{equation}
\frac{N_{\text {excited }}}{N_{\text {ground }}}=\frac{n_{\mathrm{e}} k_{01}}{n_{\mathrm{e}} k_{10}+A_{10}},
\label{eq: density}
\end{equation}
Here, ${k_\mathrm{01}}$ and ${k_\mathrm{10}}$ are the upward and downward collisional rate coefficients, respectively, and $\mathrm{A_{10}}$ is the spontaneous decay rate. We then calculated the density using the \textit{N}(\ion{C}{2}*)/\textit{N}(\ion{C}{2}) and \textit{N}(\ion{Si}{2}*)/\textit{N}(\ion{Si}{2}) ratios following \cite{Qu19}.

The level populations at different temperatures and densities are calculated with the atomic data from CHIANTI 10.0.2 \citep{Dere97, Del21}. Since both ratios are insensitive to temperature across all density ranges, we adopt a typical temperature of 10$^4$ K for estimating n$_{\mathrm{e}}$. As shown in Fig. \ref{fig: SiII_ratio}, the ratio of \textit{N}(\ion{C}{2}*)/\textit{N}(\ion{C}{2}) have large uncertainties, and the ratio becomes nearly constant when the density exceeds $10^{2.2} \ \mathrm{cm}^{-3}$. Therefore, we use \textit{N}(\ion{Si}{2}*)/\textit{N}(\ion{Si}{2}) to estimate $\mathrm{n_e}$ and find that the $\mathrm{n_e}$ of all components are above $10^{3} \ \mathrm{cm}^{-3}$, consistent with the lower limits inferred from the \ion{C}{2} ratio. For component \textbf{b}, the \ion{Si}{2} does not have enough unmasked pixels to yield an accurate column density, we adopt the lower limit of $n_{\mathrm{e}}$ derived from the \ion{C}{2} ratio, which is $>$ 2304 cm$^{-3}$. 

We also derive the n$_{\mathrm{e}}$ as a single component using the column density ratios obtained from the AOD method. The results are indicated by black dots in Fig. \ref{fig: SiII_ratio}. The electron number density using AOD method derived from the \ion{C}{2} ratio is approximately one order of magnitude lower than that derived from \ion{Si}{2}. When comparing n$_{\mathrm{e}}$ values derived from \ion{C}{2} and \ion{Si}{2} individually, those obtained via the AOD method are consistently lower than the results from decomposition fitting.

\subsection{Photoionization Analysis}
\label{sec: pie}
We use the spectral synthesis code \texttt{Cloudy} (version c17.01, \citealt{Ferland13, Ferland17}) to determine the hydrogen column density $N_\mathrm{H}$ and ionization parameter $U_\mathrm{H}$. Following the approach used in previous studies \citep{arav13, Xu18, Xu19}, we employ \texttt{Cloudy} to generate a grid of simulated models corresponding to different values of $N_\mathrm{H}$ and $U_\mathrm{H}$, assuming solar metallicity and utilizing the spectral energy distribution (SED) of a radio-quiet quasar HE 0238-1904 (hereafter HE0238; \citealt{arav13}). For the observed data in the 835-1635 \AA \ rest frame, the ratio between the HE0238 SED and the J014741 continuum remains constant within $\pm 10 \%$. The electron densities derived in Sec. \ref{sec: density} are adopted as input parameters for these simulations. The best-fit total hydrogen column density $N_H$ and ionization parameter $U_H$ are then obtained by minimizing the $\chi^2$. All solutions of different components are summerized in Fig. \ref{fig: different_solution}.

\begin{figure}
\includegraphics[width=1.0\columnwidth]{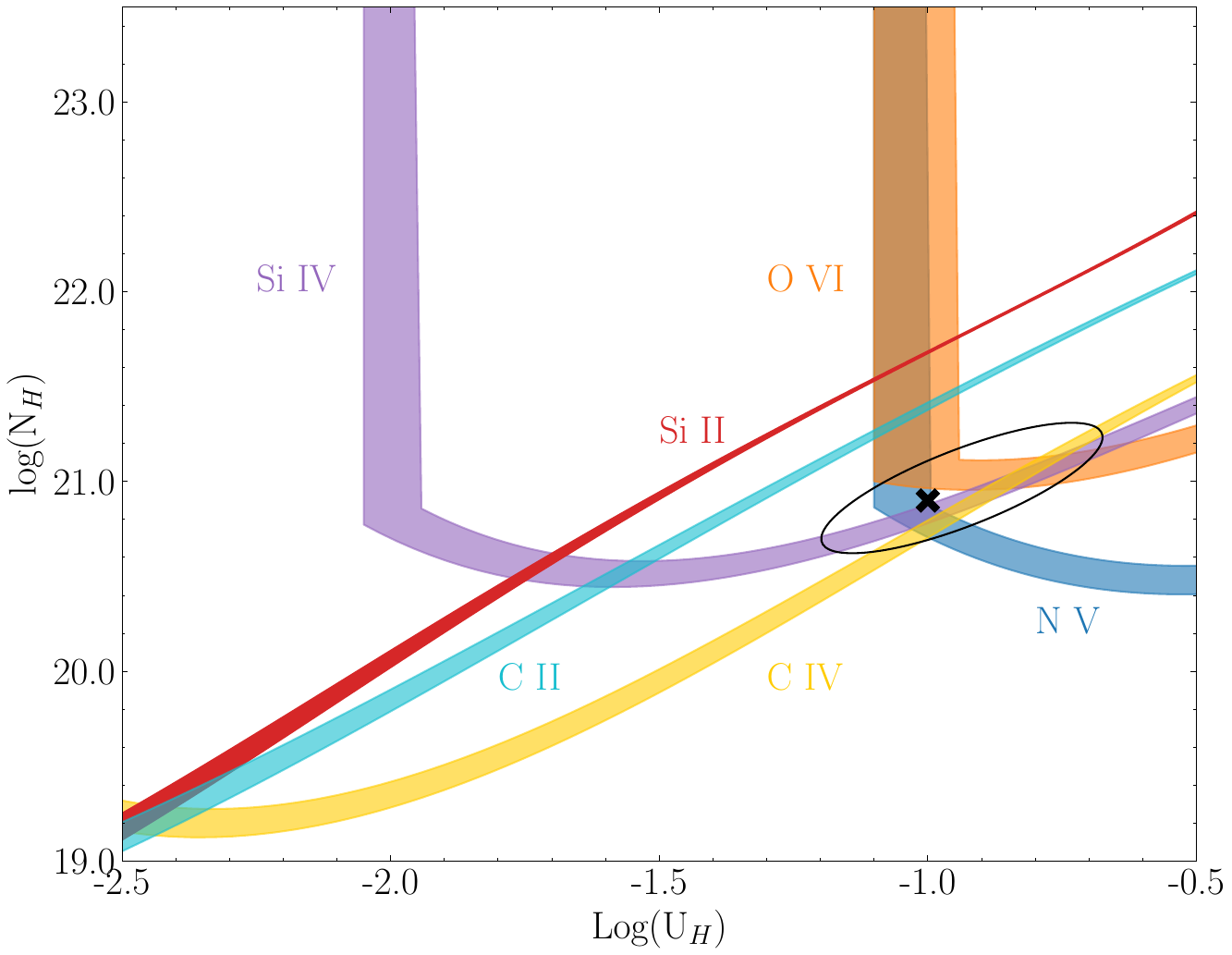}
\caption{Grid-models showing the photoionization solution of the component \textbf{f}, assuming a hydrogen density of 957 cm$^{-3}$ estimated by N(\ion{Si}{2}*)/N(\ion{Si}{2}). The range of possible solutions of different ions is labeled with different colors. The best-fit solution is marked with a black `X' with a one-sigma contour.
\label{fig: solution_f}}
\end{figure}

For each component, we compute a one-phase photoionization solution. For the majority of components, such a one-phase solution can satisfactorily reproduce the observed column densities of different ions. However, in certain cases, such as component \textbf{f} (shown in Fig. \ref{fig: solution_f}), where absorption is strong across a wide ionization range from \ion{C}{2} to \ion{O}{6}, the one-phase solution appears insufficient to account for the measured column densities of low ions like \ion{C}{2} and \ion{Si}{2}. If additional low ions such as \ion{Fe}{2} or \ion{O}{1} were detected, this would further support the need for an additional phase, contributing to a solution with a lower $U_H$. The results for other components are shown in Appendix.

If the absorber is treated as a single component and the column densities of various ions are derived using the AOD method, the resulting one-phase photoionization solution fails to simultaneously reproduce the observed column densities of all ions. On one hand, the AOD method provides only lower limits to the column densities, often accompanied by significant uncertainties. On the other hand, when the absorption profile is decomposed into individual components, the derived one-phase solutions vary significantly across components, with differences in log($U_\mathrm{H}$) reaching to $\sim$ 1.5 dex. When the entire absorption trough is modeled as a single component, the column densities from all sub-components are summed together. As a result, the combined ion column density becomes difficult to reconcile with a single ionization phase, indicating that a one-phase solution cannot adequately reproduce the total observed features.

\begin{figure}
\includegraphics[width=1.0\columnwidth]{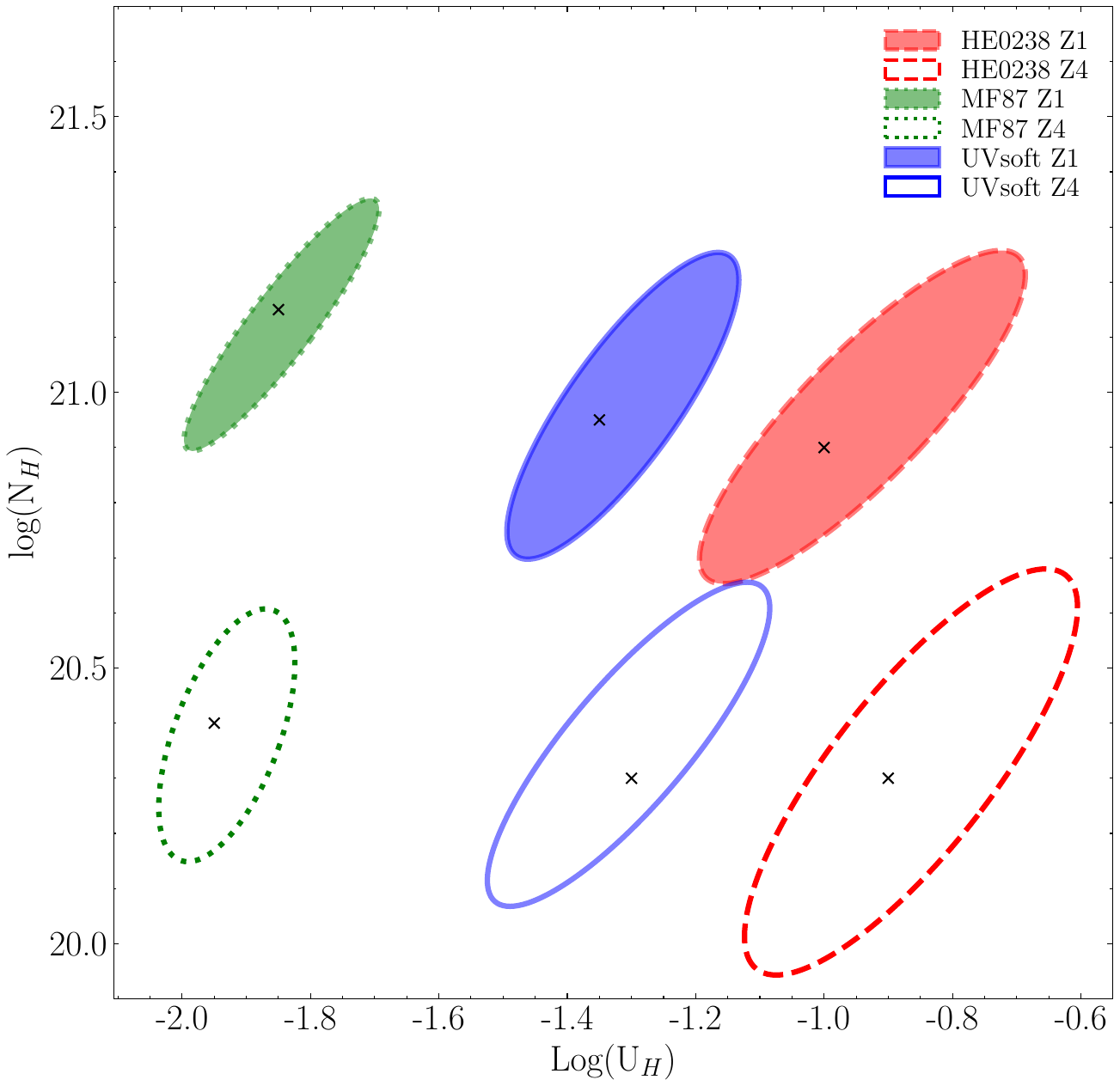}
\caption{Grid-models displaying the photoionization solutions of component \textbf{f} for three SEDs (HE0238, MF87, and UV-soft) and two metallicities: solar metallicity (Z1) and four times solar metallicity (Z4).
\label{fig: SED}}
\end{figure}

To study the influence of different SEDs and metalicities on the photoionization solutions, we consider three different SEDs and metallicities following the method of previous work (e.g., \citealt{arav13, Chamberlain15, Xu18}). In addition to the SED of HE0238, we also test the SED MF87 representative of radio-loud quasars \citep{Mathews87} and SED UV-soft from ratio-quiet quasars \citep{Dun10}. Taking component \textbf{f} as an example (shown in Fig. \ref{fig: SED}), we find that higher metallicity reduces the derived values of $\mathrm{logN_H}$. For instance, increasing the metallicity to four times the solar value, using abundance ratios from \citep{Ballero08}, results in a total hydrogen column density that is approximately 0.7 dex lower than that obtained under solar metallicity. The choice of different SEDs primarily affects the derived ionization parameters. When the SED is chosen as UV-soft, the Log(U$_H$) is approximately 0.4 smaller than when the SED is HE0238. However, when the SED is MF87, it is about 0.9 smaller than HE0238. While using the other two SED, the estimated distance will be larger, resulting in a larger mass flow rate and kinetic luminosity.

\section{Discussion} \label{sec: discuss}

As mentioned in Sec. \ref{sec: data}, the redshift of the quasar was determined based on broad emission lines, which may lead to an underestimation due to their potential blueshifts. We do not detect any narrow emission lines in the near-infrared P200/TripleSpec spectrum of this quasar, making it difficult to determine an accurate redshift\footnote{Thanks to Jingyi Yang for providing the data and data reduction}. If further infrared observations at longer wavelengths detect the H$\beta$ or [\ion{O}{3}] lines, we may be able to determine the redshift with greater precision. Since both the mass flow rate and the kinetic luminosity depend on the velocity, the quasar redshift has a significant impact on the estimation of these two quantities. The absorptions of S2 and S3 are clearly located on the red side of the emission line. Therefore, we adopt the currently measured redshift ($z \sim 4.75$) as a lower limit and the redshift of S3 ($z \sim 4.836$) as the upper limit to estimate the physical properties of the gas.

\subsection{Covering Fractions of Different Ions}
\label{sec: cf}
\begin{figure}
\includegraphics[width=1.0\columnwidth]{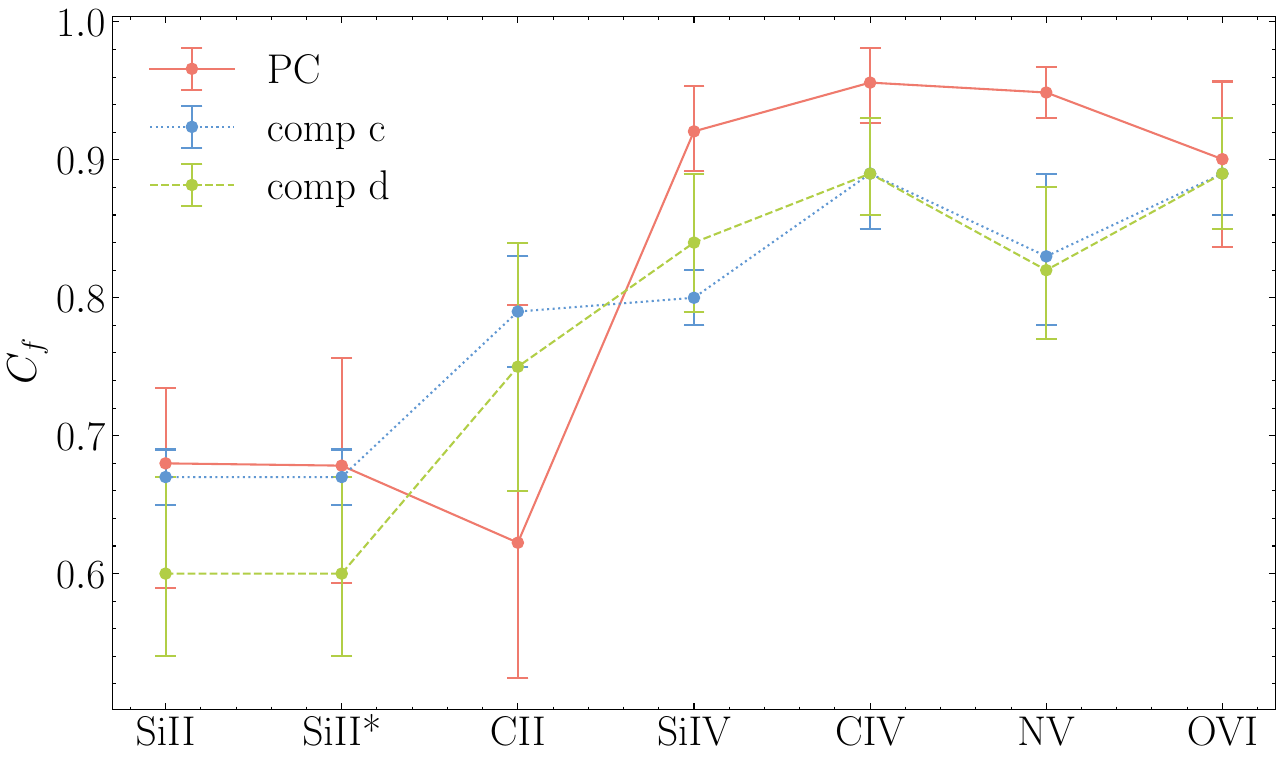}
\caption{Comparison of covering fractions among different ions. The ions are arranged in order of increasing ionization potential. Red dots represent covering fractions estimated using the Partial Covering method within  $|\Delta v| < 60~\rm km~s^{-1}$ for the S1 system. The blue and green dots represent the covering fractions of components \textbf{c} and \textbf{d}, respectively, measured via Voigt profile fitting. 
\label{fig: cf_compare}}
\end{figure}

Although all of the detected lines in S1 are optically thick, none reach zero intensity in our spectra, indicating partial covering by gas that does not fully obscure the background source. In Sec. \ref{sec: PC}, we use the PC method to calculate the covering fractions of different ions, and the results are presented in Fig. \ref{fig:partial_covering}. We mask regions contaminated by airglow emission lines and other ions, which vary among different ions, leading to differences in the number and positions of velocity bins available for calculation. Here we use the PC method to calculate the mean covering fractions between -60 km s$^{-1}$ and 60 km s$^{-1}$, as most absorption lines remain unmasked in this range. In the decomposition fitting, absorption within this range is primarily contributed by components \textbf{c} and \textbf{d}. Therefore, we also compare the covering fractions of these two components with the results obtained from the PC method. A comparison of covering fractions among different ions and different methods is shown in Fig. \ref{fig: cf_compare}. 

The \ion{C}{4} and \ion{Si}{4} absorption features lie on top of strong broad emission lines (BELs), suggesting that the absorbers only partially cover the extended BEL region. The typical size of such absorbers could be comparable to that of broad-line region (0.1-1 pc, e.g., \citealt{Bentz13}). However, the partial coverage observed in \ion{C}{2} and \ion{Si}{2} lines implies that the corresponding size of the clouds may be smaller than the continuum source, whose typical size is less than 0.01 pc \citep{Morgan10}. This conclusion is consistent with the path length size calculated from $N_{\mathrm{H}} / n_{\mathrm e}$ in our case, which is from 0.01 pc to 0.1 pc.

The significant variations in covering fractions among different ions provides strong evidence for an inhomogeneous absorbing structure \citep{Hamann97, Telfer98, arav99b, Gabel03, hamann19}. Despite differences in method, the covering fractions consistently increase with ionization potential, suggesting that high ions are more spatially extended than low ions in our case. For component \textbf{c} and \textbf{d}, which generate strong absorption in all ions, the result supports a simple geometric model in which a dense core or flow tube of low ions is surrounded by a more diffuse, high ions medium. 

\begin{figure}
\includegraphics[width=1.0\columnwidth]{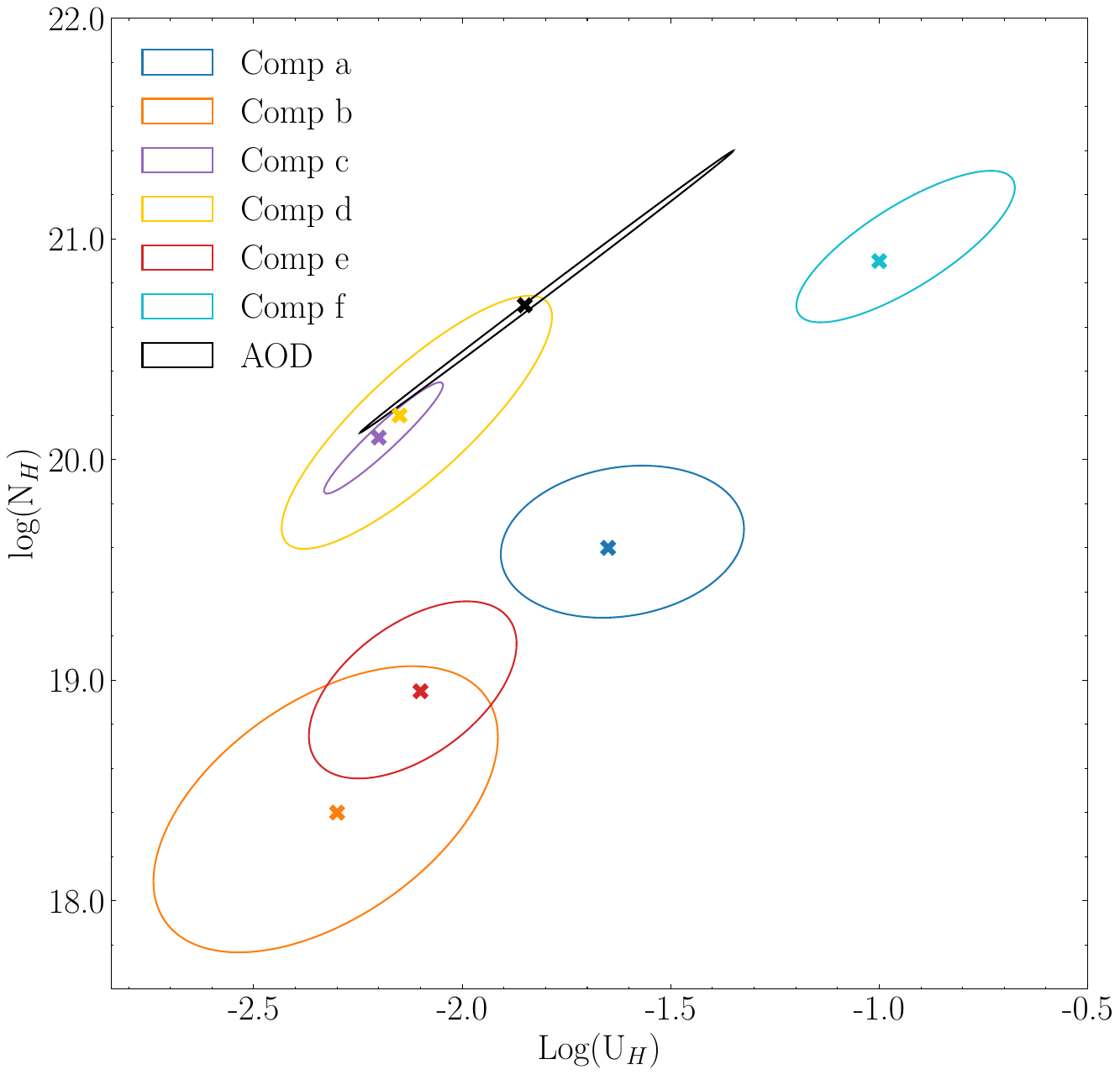}
\caption{The comparison of photoionization solutions between different components and the AOD method. The `X' markers indicate the solutions found via $\chi^2$ analysis, the elllipses represent the 1 $\sigma$ error.
\label{fig: different_solution}}
\end{figure}

\subsection{Distance Estimation}
The distance of the gas from the quasar could be derived using this relationship. The definition for the ionization parameter is listed below, 
\begin{equation}
U_{\mathrm{H}}=\frac{Q_{\mathrm{H}}}{4 \pi R^2 n_{\mathrm{H}} c},
\label{eq: ufactor}
\end{equation}
Where $Q_\mathrm{H}$ is the source emission rate of hydrogen ionizing photons, $R$ is the distance of the outflow to the central source, $n_\mathrm{H}$ is the number density of hydrogen, and $c$ is the speed of light. In high ions, the hydrogen number density $n_{\mathrm{H}}$ can be estimated using the relation $n_{\mathrm{e}} \simeq 1.2n_{\mathrm{H}}$. The hydrogen-ionizing photon rate $Q_{\mathrm{H}}$ is obtained by integrating the quasar’s SED over energies greater than 13.6 eV. We use photometric data from the Two Micron All Sky Survey (2MASS; \citealt{Skrutskie06}) to get the bolometric luminosity for rescaling the SED. This quasar is firmly detected in all bands in 2MASS, at $\mathrm{J_{Vega}}$ = 16.91 $\pm$ 0.15, $\mathrm{H_{Vega}}$ = 16.63 $\pm$ 0.26, $\mathrm{K_{Vega}}$ = 15.52 $\pm$ 0.01, corresponding to rest-frame wavelength between 2150 \AA \ and 3755 \AA. \cite{Vanden01} found that the 1300 \AA \ to 5000 \AA \ continuum roughly has a power-law index $\alpha_v=-0.44 \pm 0.10$. Using this equation, we estimate the optical luminosity at rest-frame 3000 \AA \ to be (4.77 $\pm$ 0.13) $\times$ 10$^{46}$ erg s$^{-1}$. Adopting the empirical factor \citep{Shen11} to convert the luminosity at 3000 \AA \ to the bolometric luminosity gives $L_{\mathrm{bol}}=(2.6 \pm 0.7) \times 10^{47} \mathrm{erg} \ \mathrm{s}^{-1}$. The emission rate of hydrogen-ionizing photons ($Q_\mathrm{H}$) is $Q_\mathrm{H} = (1.9\pm0.5) \times 10^{57} \ \mathrm{s}^{-1}$ by scaling the HE0238 SED to match the bolometric luminosity. 

\begin{figure}
\includegraphics[width=1.0\columnwidth]{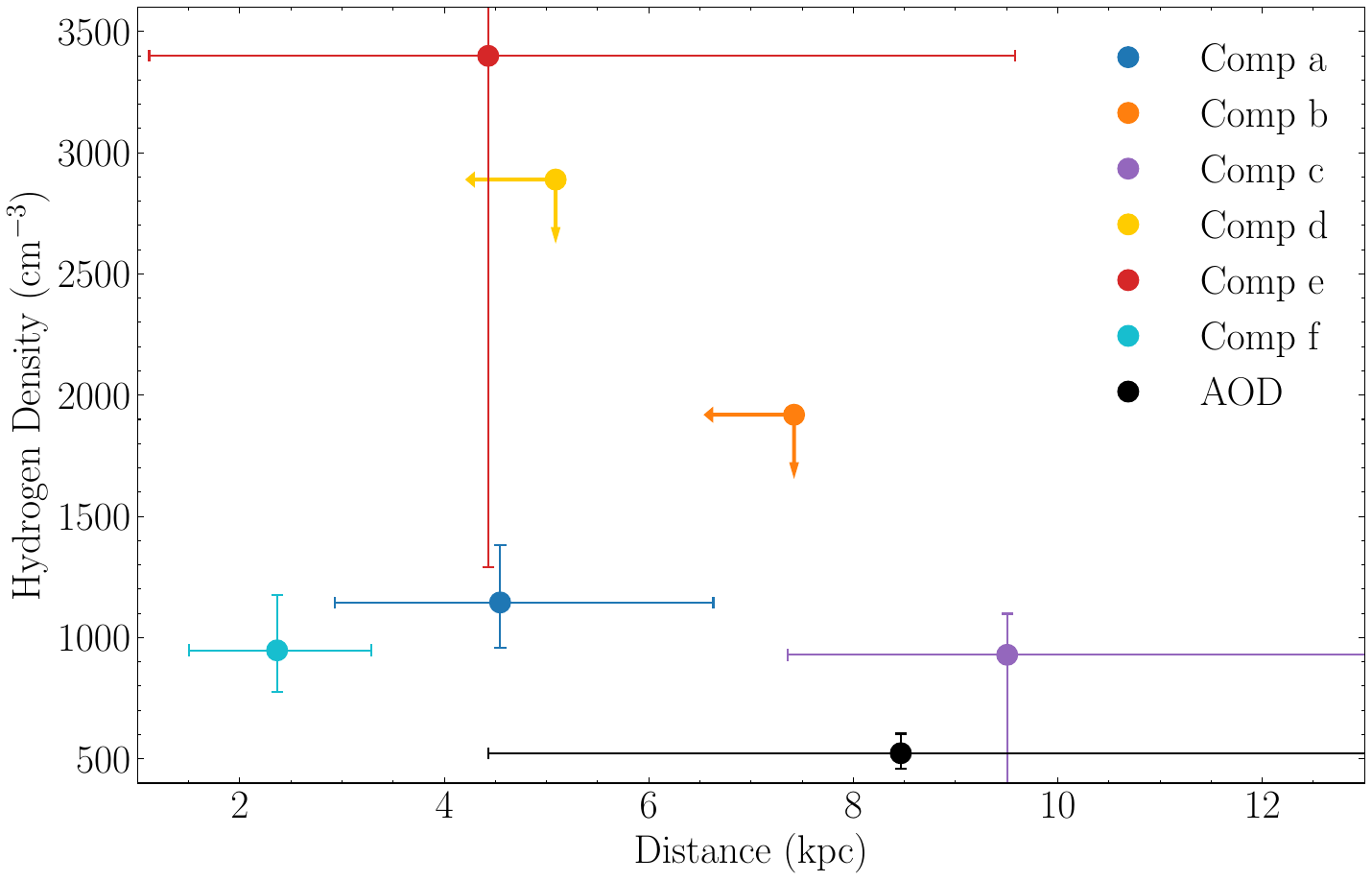}
\caption{The distance of the gas from the quasar vs. hydrogen density. Results from different components are marked with different colors, and the result from the AOD method is marked with black. The distance of component \textbf{b} is calculated with the density lower limit got from N(\ion{C}{2}*)/N(\ion{C}{2}). Since we cannot get the upper limit of the density of component \textbf{d}, the distance of this component is also presented in the form of an upper limit.
\label{fig: density_distance}}
\end{figure}

The relationship between hydrogen density and distance is plotted in Fig. \ref{fig: density_distance}. In general, the distance of the gas from the quasar is on galaxy scales, typically between 1 and 10 kpc. The size of absorbers is smaller than 1 pc even 0.01 pc because of the partial covering of quasar continuum source (Sec. \ref{sec: cf}). While such clouds are rare in the normal circumgalactic medium, they are not unprecedented in distant AAL absorbers. \cite{Hamann01, Chen18} found partial covering of the continuum source in an AAL absorber $\sim$ 28 kpc and $\sim$ 5.7 kpc from the central quasar separately. \cite{Chen18} inferred that the clouds were more likely created \textit{in situ}, at or near their observed location rather than originated from the central region of the black hole. The sound crossing time for a single cloud is defined as,
\begin{equation}
t_{\mathrm{sc}}=\frac{l}{c_{\mathrm{s}}},
\end{equation}
where $c_{\mathrm{s}}$ is the sound speed and $l$ is the characteristic cloud size. If not confined by an external pressure, clouds are expected to dissipate due to instabilities within one to several tens of sound crossing times (e.g., \citealt{Klein94, Li20, Gronke22}). For a nominal temperature of 10$^4$ K and $l \lesssim 1$ pc for conservation, the cloud survival time is estimated to be $\lesssim$ 7 $\times$ 10$^4$ yr. The mean distance of the components from the quasar is about 5 kpc. If interpreted as outflows ($z_{\mathrm{QSO}} \sim 4.836$), the corresponding flow time is $\sim$ 1.2 $\times$ 10$^6$ yr, which is about 17 times longer than the sound crossing time. In this situation, it is difficult to distinguish whether the gas is produced \textit{in situ} or originates from the vicinity of the quasar. Given the high number density of the cloud, we favor the interpretation that it most likely formed near the quasar.
 
We find an inverse correlation between the hydrogen density and distance across different components despite their widely varying ionization parameters. The relationship between density and distance generally follows $n(r) \propto r^{-\alpha}$ ($\alpha \sim 1-3$) under different scenarios \citep{Contopoulos17, Blandford82}. The index $\alpha$ is determined by the specific mechanism that drives the outflow, but the relationship consistently follows an inverse trend. Our results show that the densities of different components within the same system may also follow this trend.

\subsection{Mass and Kinetic Energy Outflow Rates}
 With the distance of the outflow determined, we can calculate the mass flow rate of each component \citep{Borguet12},

\begin{equation}
\dot{M} = 4 \pi \Omega R N_{\mathrm{H}} \mu m_p v,
\end{equation}

and the kinetic luminosity,
\begin{equation}
\dot{E} \simeq \frac{1}{2} \dot{M} v^2,
\end{equation}
Where R is the distance of the outflow from the central source, and $\Omega$ is the global covering factor for absorbers. $\mu$ = 1.4 is the mean atomic mass per proton, $m_\mathrm{p}$ is the proton mass, $N_H$ is the absorber's total hydrogen column density, and $v$ is the radial velocity of the outflow.

There is no way to give a direct measurement of $\Omega$, and the most commonly used method is to use the frequency at which the same type of absorber appears in quasars. \cite{Misawa07} found that the fraction of intrinsic \ion{C}{4} systems NALs is 11\%-19\%, which gives a lower limit. So we adopt 0.2 here as the value of $\Omega$. 

If calculated using the current redshift ($\sim$ 4.75), the gas from S1 would be classified as an inflow. Among the various components, the mass flow rates span three orders of magnitude (from 0.8 $\mathrm{M_{\odot}} \ \mathrm{yr}^{-1}$ to 88 $\mathrm{M}_{\odot} \ \mathrm{yr}^{-1}$), with component \textbf{f} overwhelmingly dominating the total outflow. The mass flow rate estimated from the column density derived using the AOD method is slightly higher than the total mass flow rate summed over the six components. 

The highest redshift of detected absorbers in S3 is 4.836, If this redshift is taken as quasar's redshift, then the absorber would be an outflow. We then recalculate the mass flow rate and derive the kinetic luminosity of the gas to evaluate outflow's feedback on the host galaxy. Theoretical models and simulations show that outflows carrying $\sim$ 5\% of the quasar's luminosity can disrupt and sweep up the cold, dense ISM that dominates the total gas mass \citep{Scannapieco04}. In the case of `multistage' model, if feedback drives a weak outflow in the hot, diffuse ISM, the resulting instabilities and pressure gradients can cause nearby cold clouds to expand perpendicular to the flow, reducing their density and potentially suppressing star formation \citep{Hopkins10}. Only 0.1\% - 0.5\% of the quasar's luminosity is required as the energy input from the outflowing gas in this scenario. The kinetic luminosity obtained from the AOD method compared to the bolometric luminosity ($\dot{E}_{\mathrm{k}}$ / $L_{\mathrm{bol}}$) is about 0.4\%. The most energetic component, component \textbf{f}, contributes only 0.15\% of $L_{\mathrm{bol}}$, suggesting that only under the `multistage' model can the outflow in our case have sufficient impact on the cold gas to act as a trigger. All physical properties including distances, mass flow rates, and energetic luminosities of different components are listed in Table \ref{table: result}.

\begin{figure}
\includegraphics[width=1.0\columnwidth]{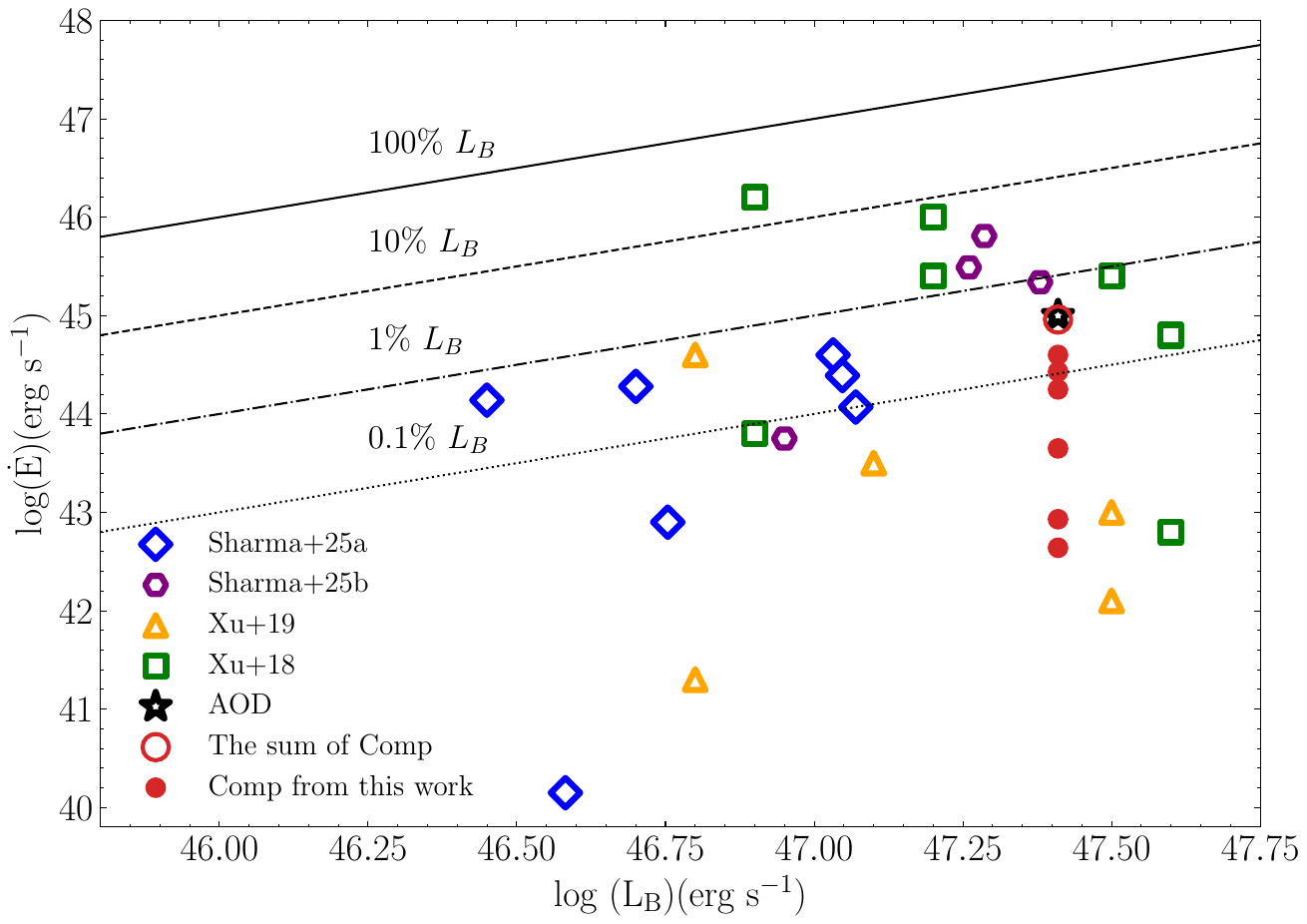}
\caption{The kinetic luminosity of gas flow from this work and other samples. $\dot{E}$ from other work is also calculated from quasar absorption lines \citep{Xu18, Xu19,Sharma25a, Sharma25b}.}
\label{fig: Sample_compare}
\end{figure}

We compare the kinetic luminosity of the outflow with that of luminous quasars ($L_{\mathrm{bol}} > 10^{46.5}$ erg/s) whose luminosities were derived using the same method. The results are shown in Fig. \ref{fig: Sample_compare}. Overall, outflows from luminous quasars tend to have limited impact on their host galaxies. The total kinetic luminositiy of all components, as well as the value derived from the AOD method, does not differ significantly from the those obtained in other cases. Two components exceed the threshold of 0.1\% $L_{\mathrm{B}}$, and even the weakest component remains comparable to some weak outflows observed in other cases.


\begin{table*}
\begin{center}
\caption{Physical parameters of different components from S1 assume that $z=$4.75.}

\hspace{-1in}
\tabcolsep=3.5pt%
\begin{tabular}{lcccccccc}
\hline\hline
Object & log($U_{H}$) & log($N_{H}$) & log($n_e$) & log(R) & $\dot{M}$ & log$\dot{E_{k}}$ & $\dot{E}_k / L_{\text {Edd }}^d$ \\
 & (km s$^{-1})$ & (log(cm$^{-2}$)) & (log(cm$^{-3}$)) & (log(pc)) & $\left(M_{\odot} \mathrm{yr}^{-1}\right)$ &  (log(erg s$^{-1}$)) & (\%) \\

\hline
comp a 
 & -1.65$^{+0.30}_{-0.25}$ & 19.60$^{+0.35}_{-0.30}$ & 3.14$^{+0.09}_{-0.08}$ & 3.66$^{+0.12}_{-0.23}$ & 7.7$^{+15.3}_{-5.4}$ & 42.73$^{+0.47}_{-0.53}$ & 0.002$^{+0.005}_{-0.001}$\\
comp b 
 & -2.30$^{+0.35}_{-0.40}$ & 18.40$^{+0.65}_{-0.65}$ & $>$3.36 & $<$3.87  & $<$0.8 & $<$41.78 & $<$0.0002\\
comp c 
 & -2.20$^{+0.15}_{-0.15}$ & 20.10$^{+0.25}_{-0.25}$ & 3.05$^{+0.07}_{-0.69}$ & 3.98$^{+0.07}_{-0.15}$ & 53.8$^{+58.6}_{-32.4}$ & 43.62$^{+0.32}_{-0.40}$ & 0.016$^{+0.059}_{-0.009}$\\
comp d
 & -2.15$^{+0.35}_{-0.30}$ & 20.20$^{+0.50}_{-0.60}$ & $>$3.54 & $<$3.71 & $<$36.8 & $<$43.47 & $<$0.011\\
comp e
 & -2.10$^{+0.20}_{-0.25}$ & 18.95$^{+0.40}_{-0.40}$ & 3.61$^{+1.00}_{-0.42}$ & 3.65$^{+0.30}_{-0.64}$ & 1.8$^{+7.3}_{-1.7}$ & 42.18$^{+0.70}_{-1.04}$ & 0.001$^{+0.003}_{-0.001}$\\
comp f
 & -1.00$^{+0.30}_{-0.20}$ & 20.90$^{+0.40}_{-0.25}$ & 3.06$^{+0.09}_{-0.09}$ & 3.37$^{+0.10}_{-0.24}$ & 88.0$^{+192.7}_{-59.3}$ & 43.87$^{+0.50}_{-0.49}$ & 0.029$^{+0.071}_{-0.018}$\\
AOD
 & -1.85$^{+0.50}_{-0.40}$ & 20.70$^{+0.70}_{-0.55}$ & 2.80$^{+0.06}_{-0.06}$ & 3.93$^{+0.23}_{-0.32}$ & 186.2$^{+1396.7}_{-160.8}$ & 44.13$^{+0.93}_{-0.87}$ & 0.053$^{+0.398}_{-0.045}$\\

\hline \hline

\end{tabular}\label{table: result}

\footnotesize

\end{center}
\end{table*}

\begin{table}

\caption{Velocity dependency parameters of different components from S1 assume that $z=$4.836.}

\hspace{-1in}
\tabcolsep=3.5pt%
\begin{tabular}{lcccccccc}
\hline\hline
Object & $\dot{M}$ & log$\dot{E_{k}}$ & $\dot{E}_k / L_{\text {Edd }}^d$ \\
 & $\left(M_{\odot} \mathrm{yr}^{-1}\right)$ &  (log(erg s$^{-1}$)) & (\%) \\

\hline
comp a 
 & 15.6$^{+35.5}_{-10.6}$ & 43.65$^{+0.51}_{-0.49}$ & 0.017$^{+0.039}_{-0.012}$\\
comp b 
 & $<$1.6 & $<$42.64 & $<$0.002\\
comp c 
 & 100.6$^{+368.4}_{-56.8}$ & 44.43$^{+0.67}_{-0.36}$ & 0.105$^{+0.385}_{-0.059}$\\
comp d
 & $<$67.2 & $<$44.25 & $<$0.069\\
comp e
 & 3.3$^{+14.5}_{-2.9}$ & 42.93$^{+0.74}_{-1.0}$ & 0.003$^{+0.015}_{-0.006}$\\
comp f
 & 154.4$^{+384.5}_{-99.2}$ & 44.6$^{+0.54}_{-0.45}$ & 0.154$^{+0.385}_{-0.099}$\\
AOD
 & 361.8$^{+2710.5}_{-308.4}$ & 45.0$^{+0.93}_{-0.83}$ & 0.390$^{+2.919}_{-0.332}$\\

\hline \hline

\end{tabular}\label{table: result_z2}

\footnotesize

\end{table}

\section{Summary} \label{sec: summary}
We present a component by component analysis on three absorption systems of a quasar J014741 ($z\sim 4.75$) based on the high-resolution spectrum from Magellan/MIKE. Our main results are as follows:

1. We find three absorption systems (S1-S3) spanning the velocity range from 1600 km s$^{-1}$ to 4500 km s$^{-1}$ ($z_\mathrm{S1} \sim 4.780$, $z_\mathrm{S2} \sim 4.829$, $z_\mathrm{S3} \sim 4.836$). We identify various absorption troughs from both highly and low ions. Then we use the AOD method and decomposition fitting to get the column densities separately. The results show that AOD method underestimates the column densities especially for the high ions.

2. In S1, density-sensitive ratios \textit{N}(\ion{Si}{2}*)/\textit{N}(\ion{Si}{2}) and  \textit{N}(\ion{C}{2}*)/\textit{N}(\ion{C}{2}) determine the electron number density. Both ratios indicate that the gas is dense and that the electron number density is greater than 10$^{2.8}$ cm$^{-3}$ for all components. 

3. With the $n_e$ determined, we use the derived column densities to find the best solution in the $\log \left(N_{\mathrm{H}}\right)-\log \left(U_{\mathrm{H}}\right)$ phase space for each component in the first system. We choose the HE0234 SED and solar metallicity as the input value and test the dependency of the solution on the choice of SED and metallicities. Then we compare the solutions of the AOD method with the decomposing method. The results show that the total hydrogen column density derived from the AOD method is close to the largest component value and its ionization parameter lies in the range of all components. 

4. Both the PC method and decomposition fitting for S1 indicate a partial coverage of the gas. This indicates that the absorbers sizes are $\lesssim$ 1 pc or even $\lesssim$ 0.01 pc (based on \ion{C}{2} and \ion{Si}{2} that partially cover the quasar continuum source). We also find that ions in higher ionized states tend to have larger covering fractions, which means that the gas in higher ionized states may have a more extended distribution compared to the gas in lower ionized states.

5. Combining $n_e$ derived from the line ratios, we infer that the gas is located at a distance of 2.3-9.5 kpc from the quasar. The flow time of the gas is about 17 times longer than its sound crossing time, making it difficult to constrain its origin. Nevertheless, the high number density may be more consistent with an origin in the vicinity of the quasar.

6. Since redshifts derived from broad emission lines may underestimate the true redshift of the quasar, we also adopt the redshift of the absorption lines as an upper limit for the quasar redshift. Using both redshift estimates, we separately calculate the gas mass flow rate and kinetic luminosity. In both cases, the kinetic luminosity ($< 0.5\% L_{B}$) is too low to drive significant AGN feedback and may only suppress star formation through a `multistage' model. 

\begin{acknowledgements}
J.T.L. acknowledges the financial support from the China Manned Space Program with grant no. CMS-CSST-2025-A04 and CMS-CSST-2025-A10, and the National Science Foundation of China (NSFC) through the grants 12321003 and 12273111.
\end{acknowledgements}

\appendix
\section{The solution of each component}
\begin{figure}
\includegraphics[width=1.0\columnwidth]{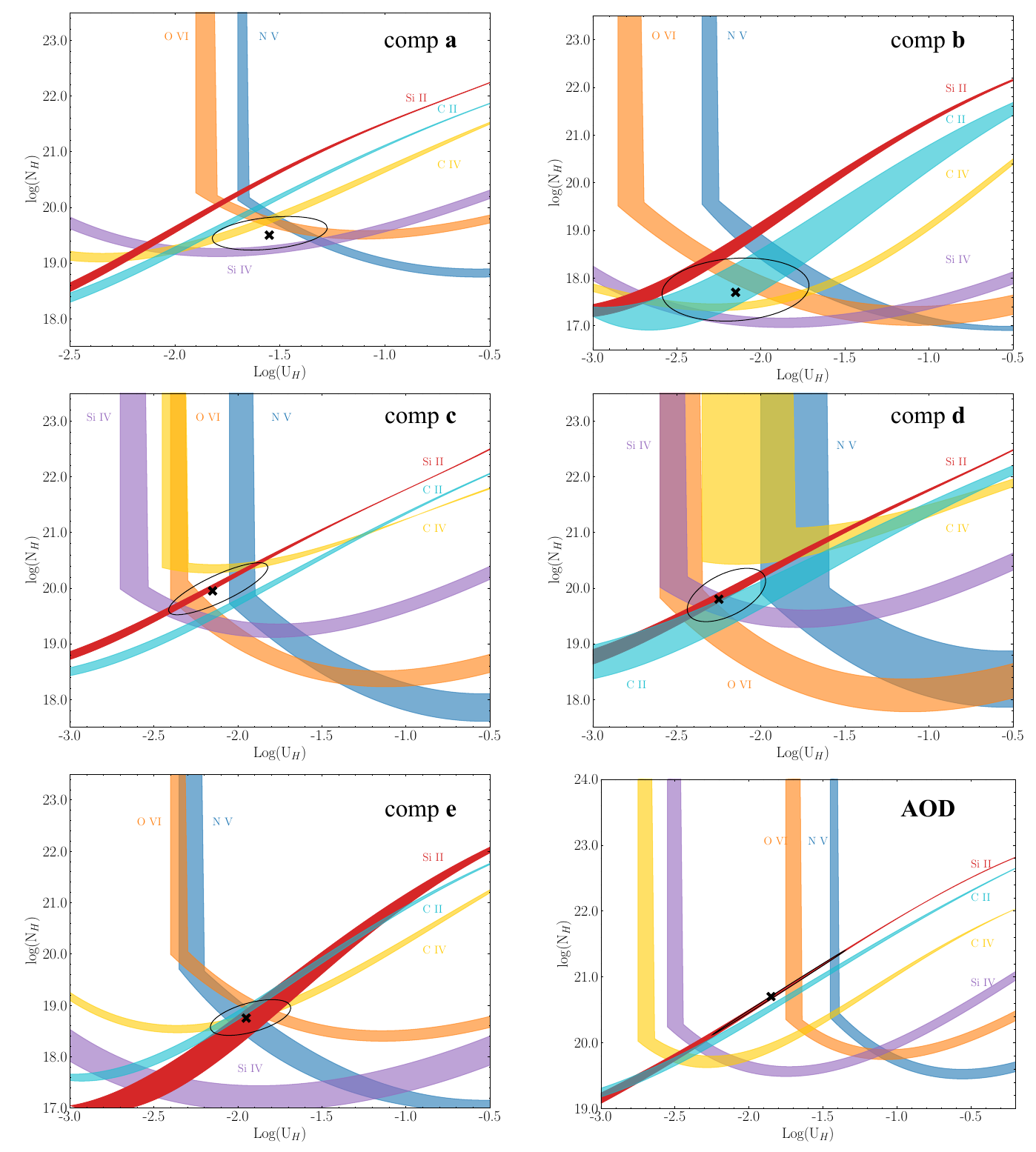}
\caption{Plots of log$N_{\mathrm{H}}$ versus log$U_{\mathrm{H}}$ for different components. This figure presents the detailed solutions for all components, serving as an expanded view of Fig. \ref{fig: different_solution}. As in Fig. \ref{fig: component_solution}, the colored shaded regions mark the solution ranges constrained by different ions. The black `X' indicates the best-fit solution and the black ellipse represents the 1 $\sigma$ error. 
\label{fig: component_solution}}
\end{figure}

In the main text, the photoionization solutions of all components are summarized in Fig. \ref{fig: different_solution}, with a detailed solution for component \textbf{f} shown in Fig. \ref{fig: component_solution}. For completeness, the full set of solutions for the remaining components is provided in this appendix for reference.

\bibliography{sample631}{}
\bibliographystyle{aasjournal}

\end{document}